\documentclass[%
 aip,
 amsmath,amssymb,
 reprint,%
]{revtex4-2}

\usepackage{graphicx, caption}
\usepackage{dcolumn}
\usepackage{bm}

\usepackage[utf8]{inputenc}
\usepackage[T1]{fontenc}
\usepackage{mathptmx}

\usepackage{threeparttable }

\usepackage{cleveref}

\usepackage{enumitem,kantlipsum}
\usepackage{etoolbox}
\usepackage[version=4]{mhchem}
\usepackage{xcolor}
\usepackage{float}
\usepackage{appendix}
\usepackage{adjustbox}
\usepackage{wrapfig}

\usepackage[figuresright]{rotating}

\Crefname{equation}{Eq.}{Eqs.}

\makeatletter
\def\@email#1#2{%
 \endgroup
 \patchcmd{\titleblock@produce}
  {\frontmatter@RRAPformat}
  {\frontmatter@RRAPformat{\produce@RRAP{*#1\href{mailto:#2}{#2}}}\frontmatter@RRAPformat}
  {}{}
}%
\makeatother

\begin{document}

\preprint{AIP/123-QED}

\title{Multiconfigurational short-range on-top pair-density functional theory}

\author{Frederik Kamper Jørgensen}
\affiliation{Department of Physics, Chemistry and Pharmacy, University of Southern Denmark, Campusvej~55, DK--5230 Odense M, Denmark}

\author{Erik Rosendahl Kjellgren}
\affiliation{Department of Physics, Chemistry and Pharmacy, University of Southern Denmark, Campusvej~55, DK--5230 Odense M, Denmark}

\author{Hans Jørgen Aagaard Jensen}
\affiliation{Department of Physics, Chemistry and Pharmacy, University of Southern Denmark, Campusvej~55, DK--5230 Odense M, Denmark}
\email{hjj@sdu.dk}

\author{Erik Donovan Hedegård}
\affiliation{Department of Physics, Chemistry and Pharmacy, University of Southern Denmark, Campusvej~55, DK--5230 Odense M, Denmark}
\email{erdh@sdu.dk}

\date{\today}

\begin{abstract}
\subsection*{Abstract}\noindent
We present the theory and implementation of a fully variational wave function -- density functional theory (DFT) hybrid model, which is applicable to many cases of strong correlation. 
We denote this model the multiconfigurational self-consistent on-top pair-density functional theory model (MC-srPDFT).
We have previously shown how the multiconfigurational short-range DFT hybrid model (MC-srDFT) can describe many multiconfigurational cases of any spin symmetry, and also state-specific calculations on excited states (Hedegård \textit{et al.} \textit{J. Chem. Phys.} \textbf{148}, 2018, 214103). 
However, the srDFT part of the MC-srDFT has some deficiencies that it shares with Kohn-Sham DFT; in particular (1) self-interaction errors (albeit reduced because of the range separation), (2) that different $M_S$ states incorrectly become non-degenerate, and (3) that singlet and non-singlet states dissociating to the same open-shell fragments incorrectly lead to different electronic energies at dissociation.
The model we present in this paper corrects these deficiencies by introducing the on-top pair density as an auxiliary variable replacing the spin density. Unlike other models in the literature, our model is fully variational and employs a long-range version of the on-top pair density. The implementation is a second-order optimization algorithm ensuring robust convergence to both ground- and excited states. We show how MC-srPDFT solves the mentioned challenges by sample calculations on the ground state singlet curve of \ce{H2}, \ce{N2}, and \ce{Cr2} and the lowest triplet curves for \ce{N2} and \ce{Cr2}.
Furthermore, the rotational barrier for ethene is investigated for the $S_0$ and $T_1$ states.
The calculations show correct degeneracy between the singlet and triplet curves at dissociation and the results are invariant to the choice of $M_S$ value for the triplet curves.
\end{abstract}

\keywords{}

\maketitle

\section{INTRODUCTION}

Density functional theory in the Kohn-Sham formulation (DFT)\cite{burke2005,burke2012,jones2015,tealeDFTExchange2022} is widely used in chemistry, ranging from small/medium sized molecular systems\cite{bursch2022} over bio-molecules\cite{schlick2021} to solid states\cite{clark2005,alessandro2023}. Relativistic extensions have also emerged,\cite{saue2020,creutzberg2023} and DFT can today be employed in all parts of the periodic table. 
Yet, systems with strong correlation\cite{cohen2012,zhou2022}, i.e., systems with dense orbital manifolds, several low-lying excited states or spin correlation, are known to be problematic for DFT. These characteristics are often (but far from exclusively) met for transition metal systems, which also frequently show failures with DFT.\cite{riccardi2018,grimme2018,jorgensen2024}

One of the main obstacles for  Kohn-Sham DFT  in  strongly correlated systems is its underlying use of a single Slater determinant to construct the electron density. 
Typical examples where the single Slater determinant description fails, are bond dissociations, even for simple diatomic molecules with light atoms\cite{gunnarsson1976,seminario1994}. 
Due to spin-pairing at the equilibrium bond distance, the atomic fragments in the dissociation limit will usually display different  $M_S$ (and $S$) values than around the equilibrium distance. 
Correctly connecting the two fragment wave functions with correct $S_i, M_{S,i}$ values at the dissociation limit with the wave function with the correct  $S, M_S$ at the equilibrium distance requires several Slater determinants. 
Thus, it is an example of simple, strongly correlated system, making it an obvious test bed for new methods targeted towards strong correlation. 
Moreover, it is also a scenario for which a regular Kohn-Sham approach will result in a large deviation at the dissociation limit. 
Improvement can be obtained with unrestricted Kohn-Sham DFT, i.e.\ by allowing the breaking of the spin symmetry, which necessitates the inclusion of the spin density in the approximate functionals used in Kohn-Sham DFT.\cite{barth1972,rajagopal1973}
An unfortunate consequence is that it -- despite it may improve the energetics -- introduces an unphysical breaking of the spin-symmetry of the system, often causing problems for other molecular properties than the energy. 
This has been denoted as the symmetry-dilemma of DFT,\cite{perdew1995escaping}  and the same dilemma exists for Hartree–Fock theory.\cite{lowdin1955a,lowdin1955b}

Strongly correlated systems can be treated with multiconfigurational wave functions. 
These wave functions can ensure that the spin-symmetry is retained also in the dissociation limit. 
In most cases the multiconfigurational wave functions are defined through the complete active space (CAS) \textit{ansatz}. 
The CAS(\textit{m},\textit{n}) wave function includes all configurations where \textit{m} electrons are distributed in \textit{n} orbitals, i.e., a full-configuration interaction (FCI) wave function is constructed within the chosen active space. 
With an appropriately selected active space, qualitatively correct dissociation behavior is ensured. However,  unless huge active spaces are employed, CAS(\textit{m},\textit{n}) wave functions  will generally lead to too high energies around the equilibrium distance. Therefore, obtaining quantitatively correct dissociation energies can usually only be achieved through a perturbative correction. 
Examples of methods employing such corrections are CASPT2\cite{andersson1990,andersson1992} and NEVPT2.\cite{angeli2001} 
Yet, the computational cost of these methods often becomes unacceptably large. 
Thus, there is currently ongoing developments towards more efficient multiconfigurational wave functions.\cite{ghosh2018,park2020}

One route towards accurate and computationally efficient models is to combine a CAS-type wave function with DFT.\cite{jorgensen2024,ghosh2018} 
The latter is remarkably accurate, and also qualitatively correct, around the equilibrium distance. Several such methods have been formulated\cite{savin88-dft_ci,miehlich97-dft_ci,grimme1999,grafenstein00-casdft,li2014multiconfiguration,pastorczak2018,hapka2020a}. 
However, a major challenge is to avoid double-counting of electron correlation\cite{cremer2001-density}, as correlation may come from both the exchange-correlation functional and the CAS wave function.

We have in recent years developed MC-DFT hybrids\cite{fromager2007universality,hedegaard2018multiconfigurational} free of double-counting by exploiting range-separation of the two-electron operator as proposed by Savin.\cite{savin1995,savinbook} 
We denote these models in general MC \textit{short-range} DFT (MC-srDFT), and specifically CAS-srDFT if the wave function is of CAS type. 
A great advantage of these models is that \emph{all} wave function parameters, and thus also all density parameters, are simultaneously optimized, ensuring a fully variational model. 
Thereby, the model is straightforward to extend to time-dependent and time-independent properties through response theory.\cite{fromager2013,hedegaard2013,jorgensen2022} 
However, our previous open-shell MC-srDFT model\cite{hedegaard2018multiconfigurational} cannot correctly describe dissociation since the functionals are restricted to the maximum $M_S$ values within a given multiplet and no local spin (a problem they have inherited from regular KS-DFT).   

Describing multiplets correctly with the usual approximate functionals employed in Kohn-Sham DFT is another known issue, which is related to the failure to describe dissociation curves consistently. 
A solution to this issue was suggested by \citeauthor{ziegler1977calculation}\cite{ziegler1977calculation} who proposed to include a local two-electron density as a replacement for the spin density to incorporate a "correlated" description of the spin polarization. 
The quantity used was the so-called two-electron \textit{on-top pair density} which describes the probability of finding two electrons with  opposite spins at the same point in space simultaneously. 
In later papers, several authors have also advocated using the on-top pair density.\cite{moscardo91_pdft,becke1995extension} 
More recently, these methods have seen a revival in combination with multi-configurational wave functions, perhaps most prominently instigated by the group of Gagliardi\cite{li2014multiconfiguration}, although other groups have also been active.\cite{gritsenko18-caspidft,ferte2019}
The model devised by the Gagliardi group calculates the total energy as a non-variational correction, where the density and on-top pair density are extracted from an optimized CAS  wave function. 
Thus, the method is non-variational. 
However, this constraint was recently lifted by the group of Delcey.\cite{scott2024variational} 

In this work, we present a variational, second-order optimization scheme for the range-separated multiconfigurational on-top pair density hybrid. We denote this method the multiconfigurational short-range on-top pair density functional theory (MC-srPDFT). This work expands our and others' previous developments in the regime of range-separated multiconfigurational hybrids.\cite{fromager2007universality,hedegaard2013,hedegaard2018multiconfigurational,fromager2013,hapka2020a}
In this work the on-top pair density functional presented is a translated\cite{li2014multiconfiguration} version of the range-separated short-range local spin density approximation functional,\cite{paziani2006local} which we designate sr$t$LDA in accordance with the notation used in the original MC-PDFT paper.\cite{li2014multiconfiguration} 
This range-separated model was also suggested recently by Hapka \textit{et al.}\cite{hapka2020a}, but it was only derived and implemented in a non-variational form based on standard CASSCF, i.e.\ a range-separated version of CAS-PDFT.

This paper is structured as follows: In Section \ref{Theory} we derive the theory required for a second-order optimization of an MC-srPDFT wave function. 
We next showcase the method showing calculated dissociation curves for \ce{H2}, \ce{N2}, \ce{Cr2}, and the rotational barriers of singlet and triplet states of ethene. 
The computational details are given in Section \ref{ComDetails} while the results are discussed in Section \ref{Results}. 
Finally, we conclude the paper in Section \ref{Conclusion}. 

\section{THEORETICAL BACKGROUND}\label{Theory}
In this Section the theory of the multiconfigurational short-range on-top pair-density model is presented.
We first briefly summarize the generic multiconfigurational short-range density functional theory (MC-srDFT) model and the underlying equations for the construction of charge- and spin densities from the long-range wave function. 
These quantities are essential for the short-range exchange-correlation functionals.
Next we recap how the one-electron spin-density can be replaced by the two-electron on-top pair density, and we examine how this alters the working equations for a direct second-order wave function optimization. 
We generally work in atomic units throughout this Section. 

\subsection{The MC-srDFT model}\label{Theory_mcsrdft}
In the range-separated hybrid MC-srDFT model, the electronic energy is separable as the sum of a long-range and a short-range contribution
\begin{align}\label{mcsrdft_energy}
    E(\boldsymbol{\lambda})=E^\text{lr}(\boldsymbol{\lambda})+E^\text{sr}(\boldsymbol{\lambda}).
\end{align}
For simplicity and ease of notation, the wave function parameters are collected in the column vector $\boldsymbol{\lambda}$. 
For a multiconfigurational wave function this vector contains both the configuration coefficients, $\{\textbf{c}\}$, and the orbital rotation amplitudes, $\{\boldsymbol{\kappa}\}$, as $\boldsymbol{\lambda}^{\top}=(\textbf{c}^{\top},\boldsymbol{\kappa}^{\top})$.
The separation of the energy in Eq.~\eqref{mcsrdft_energy} is achieved by separating the two-electron repulsion term, $\quad r^{-1}_{ij}=|\textbf{r}_{i}-\textbf{r}_{j}|^{-1}$, into long-range and short-range parts,
\begin{align}
    \frac{1}{r_{ij}}=\frac{\text{erf}(\mu r_{ij})}{r_{ij}}+\frac{1-\text{erf}(\mu r_{ij})}{r_{ij}}\label{inverse_elec_repulsion}, 
\end{align}
respectively. We have here used the \emph{error function},\cite{savin1995,savinbook} where  
the range separation is controlled by the range-separation parameter $\mu$.
In exact theory the computational results would be independent of its value; however, in practice we choose an optimal value from considerations of computational efficiency together with limitations on basis set, configuration space, and available short-range functionals.
This parameter has therefore been chosen based on empirical studies\cite{hubert2016a,hubert2016b,fromager2007universality} and usually attains a value of $\mu=0.4\pm 0.1$\ bohr$^{-1}$. 
However, we note that a recent investigation on spin-spin coupling constants on transition metal complexes have hinted at a value of $\mu=1.0$ bohr$^{-1}$ to be preferable for transition metal complexes.\cite{kjellgren2021multi}

With the range separation of the two-electron term, the short-range contribution can be replaced with a density functional representation. 
This leaves the long-range contribution to be fully wave function based, as is  the case in the MCSCF method.
The MC-srDFT energy can thus be written as,
\begin{align}
    E\left(\boldsymbol{\lambda}\right)
    &=\langle \Psi^\mathrm{lr}(\boldsymbol{\lambda})|\hat{H}^{\mathrm{lr},\mu}|\Psi^\mathrm{lr}(\boldsymbol{\lambda})\rangle\nonumber\\
    &\quad+ E^{\mathrm{sr},\mu}_{\mathrm{H}}\left[\rho_\mathrm{C}(\textbf{r},\boldsymbol{\lambda})\right]+E^{\mathrm{sr},\mu}_{\mathrm{xc}}\left[\xi(\textbf{r},\boldsymbol{\lambda})\right], \label{energy_ex_H_xc}
\end{align}
where $|\Psi^\mathrm{lr}(\boldsymbol{\lambda})\rangle$ is the long-range multiconfigurational wave function (to be defined below).
The first term corresponds to $E^\text{lr}(\boldsymbol{\lambda})$ in Eq.~(\ref{mcsrdft_energy}), whereas the two last terms comprise the total short-range contribution, $E^\text{sr}(\boldsymbol{\lambda})$.
We have included the superscript $\mu$ to denote that the individual energy terms are $\mu$-dependent.

We first discuss the long-range part of the above equation. In the non-relativistic or scalar-relativistic regimes within the Born-Oppenheimer approximation, the second-quantized\cite{helgaker2004} spin-free electronic Hamiltonian becomes
\begin{align}\label{hamiltonian}
\hat{H}^{\text{lr},\mu}=\sum_{pq}h_{pq}\hat{E}_{pq}+\frac{1}{2}\sum_{pq,rs}g_{pq,rs}^{\text{lr},\mu}\hat{e}_{pq,rs}+V_{\text{NN}},
\end{align}
where the first term is the one-electron Hamiltonian with the one-electron singlet operator in second quantization $\hat{E}_{pq}=\hat{a}^\dagger_{p\alpha}\hat{a}_{q\alpha}+\hat{a}^\dagger_{p\beta}\hat{a}_{q\beta}$, the second term is the modified two-electron Hamiltonian with $\hat{e}_{pqrs}=\hat{E}_{pq}\hat{E}_{rs}-\delta_{qr}\hat{E}_{ps}$,
 and $V_{\mathrm{NN}}$ being the scalar nuclear-nuclear potential contribution.
The superscript "lr" designates the use of the long-range part of the range-separated two-electron integrals. 

In this work, the long-range multiconfigurational wave function is parameterized using an exponential unitary orbital rotation operator, configuration coefficients and a projector, $\mathcal{P}=1-|0\rangle\langle0|$, as in Ref. \citenum{helgaker2004}.
We briefly summarize the wave function expression here,
\begin{align}\label{wavefun_ansatz}
    |\Psi^\text{lr}(\textbf{c},\boldsymbol{\kappa})\rangle=\exp(-\hat{\kappa})\frac{||0\rangle+\mathcal{P}|\textbf{c}\rangle\rangle}{\sqrt{1+\langle \textbf{c}|\mathcal{P}|\textbf{c}\rangle}}, 
\end{align}
where the reference and the correction states are expressed as
\begin{align}
    |0\rangle=\sum_{i}C_{i}^{(0)}\,|i\rangle\quad\text{and}\quad|\textbf{c}\rangle=\sum_{i}c_i\,|i\rangle. 
\end{align}
In this paper, we only consider real Hamiltonians. 
Accordingly, the unitary variations within the orbital space denoted by the operator, $\hat{\kappa}$, are thus constrained to only include the special orthogonal rotations within the real orbital space
\begin{align}
    \hat{\kappa}=\sum_{p>q}\kappa_{pq}\left(\hat{E}_{pq}-\hat{E}_{qp}\right)=\sum_{p>q}\kappa_{pq}\hat{E}_{pq}^{-}, 
\end{align}
where the anti-symmetric matrix, $\boldsymbol{\kappa}$, contains the non-redundant orbital rotation parameters between the orbitals.   

Moving now to the two last terms of Eq.~\eqref{energy_ex_H_xc}, these are the two short-range functional terms, referred to as the short-range Hartree, $\Bar{E}^{\mathrm{sr},\mu}_{\mathrm{H}}\left[\rho_\mathrm{C}\right]$, and the short-range exchange-correlation, $\Bar{E}^{\mathrm{sr},\mu}_{\mathrm{xc}}\left[\xi\right]$, terms. 
The short-range functionals are complementary to the long-range energy contribution such that Eq. \eqref{mcsrdft_energy} is fulfilled. 

The Hartree term only depends on the total charge density, $\rho_\mathrm{C}$, of the molecular system.
In contrast, the exchange-correlation energy can depend on the charge density ($\rho_\mathrm{C}$), spin density ($\rho_\mathrm{S}$), and derivatives hereof. 
These possible variables entering the functional is symbolically denoted by $\xi$.
All these quantities are constructed as the expectation values of the long-range wave function. 
We illustrate this here with the charge- and spin-densities, where the corresponding operators are defined 
\begin{align}
\hat{\rho}_\text{C}(\textbf{r}) = \sum_{pq}\Omega_{pq}(\textbf{r})\hat{E}_{pq}\, \, \, \, \, \, \text{and}\, \, \, \, \, \, \, 
\hat{\rho}_\text{S}(\textbf{r}) = \sum_{pq}\Omega_{pq}(\textbf{r})\hat{T}_{pq} ,
    \label{density-operator}
\end{align}
where $\hat{E}_{pq}$ has been defined previously and $\hat{T}_{pq}=\hat{a}^\dagger_{p\alpha}\hat{a}_{q\alpha}-\hat{a}^\dagger_{p\beta}\hat{a}_{q\beta}$.
Accordingly, the expectation value becomes
\begin{align}
    \rho_\text{X}(\textbf{r},\boldsymbol{\lambda}) &= \langle \Psi^\text{lr}(\boldsymbol{\lambda})|\hat{\rho}_\text{X}(\textbf{r})|\Psi^\text{lr}(\boldsymbol{\lambda})\rangle \nonumber\\
    &= \sum_{pq}\Omega_{pq}(\textbf{r})D^\text{X}_{pq}(\boldsymbol{\lambda})\quad\text{with}\quad\text{X}\in\{\text{C},\text{S}\},
    \label{density-exp-value}
\end{align}
where $\Omega_{pq}(\textbf{r})=\phi_p^*(\textbf{r})\,\phi_q(\textbf{r})=\phi_p(\textbf{r})\,\phi_q(\textbf{r})$ is an element of the MO overlap distribution and $D^\text{X}_{pq}$ is an element of the one-electron reduced charge or spin density matrix. 
From the expressions of the singlet and triplet excitation operators, it is clear that the charge- and spin densities can be decomposed as
\begin{subequations}
\begin{align}
    \rho_\text{C} &= \rho_\alpha + \rho_\beta \\
    \rho_\text{S} &= \rho_\alpha - \rho_\beta . 
\end{align}
\end{subequations}

We finally note that the short-range Hartree functional becomes\cite{hedegaard2018multiconfigurational}
\begin{align}
    E^{\mathrm{sr},\mu}_{\mathrm{H}}\left[\rho_{C}(\textbf{r},\boldsymbol{\lambda})\right] &= \frac{1}{2}\sum_{pq,rs}D_{pq}^{\mathrm{C}}g^{\mathrm{sr},\mu}_{pq,rs}D_{rs}^\mathrm{C}=\frac{1}{2}\sum_{pq}D_{pq}^\mathrm{C} j_{pq}^{\mathrm{sr},\mu} \label{Hartree} . 
\end{align}
The short-range exchange-correlation functional is
\begin{align}
E^{\mathrm{sr},\mu}_{\mathrm{xc}}\left[\xi(\textbf{r},\boldsymbol{\lambda})\right] &=\int\text{e}^{\mathrm{sr},\mu}_\text{xc}\left(\xi(\textbf{r},\boldsymbol{\lambda})  \right)\,\text{d}\textbf{r}\label{energy_density},
\end{align}
where $\text{e}^{\mathrm{sr},\mu}_\text{xc}(\xi)$ is the short-range exchange-correlation energy density. 
In the remainder of the paper we omit the explicit $\mu$ superscripts in order not to overload the notation when we take derivatives.

\subsection{Spin-polarization through the on-top pair density}\label{Theory_ontop_pair}

The electron on-top pair-density, $\pi(\textbf{r})$, is defined as the probability of finding two electrons in the same point in space, i.e., $\textbf{r}_1=\textbf{r}_2=\textbf{r}$. 
In a range-separated approach, the on-top pair density is defined from the long-range wave function as
\begin{align}
    \pi(\textbf{r})&= \left(\begin{array}{c}
    N\\
    2
\end{array}\right)\int|\Psi^\text{lr}(\textbf{r}\sigma_1,\textbf{r}\sigma_2,\dots,\textbf{x}_N)|^2\,\nonumber\\
&\qquad\qquad\qquad\times\text{d}\sigma_1 \text{d}\sigma_2 \text{d}\textbf{x}_3,\dots,\text{d}\textbf{x}_N. 
\end{align} 
In a second-quantization formalism, the two-electron on-top pair density operator becomes
\begin{align}\label{pi_sq}
    \hat{\pi} = \sum_{pq,rs}\Omega_{pq}(\textbf{r})\Omega_{rs}(\textbf{r})\hat{e}_{pq,rs}.
\end{align}
As briefly mentioned above, in our range-separated approach, the on-top pair-density is constructed from the long-range wave function as,
\begin{align}\label{pair_density_orbs}
    \pi(\textbf{r},\boldsymbol{\lambda}) &= \langle\Psi^\mathrm{lr}(\boldsymbol{\lambda})|\hat{\pi}|\Psi^\mathrm{lr}(\boldsymbol{\lambda})\rangle \nonumber\\
    &=  \sum_{pq,rs}\Omega_{pq}(\textbf{r})\Omega_{rs}(\textbf{r})P_{pq,rs}(\boldsymbol{\lambda}) , 
\end{align}
where $P_{pq,rs}(\boldsymbol{\lambda})=\langle\Psi^\text{lr}(\boldsymbol{\lambda})|\hat{e}_{pq,rs}|\Psi^\text{lr}(\boldsymbol{\lambda})\rangle$ is the two-electron reduced density matrix. 
For a single Slater determinant wave function, the relation between the spin density and the pair density can be expressed by the squared spin-polarization factor, $\eta$,
\begin{align}\label{pi_in_spin_polari}
    \pi(\textbf{r}) = \rho_\alpha(\textbf{r})\rho_\beta(\textbf{r})=\frac{1}{4}\rho_\text{C}^2 (\textbf{r}) \left(1-\eta^2(\textbf{r})  \right),
\end{align}
which itself is defined as,
\begin{align}\label{spin_polari}
    \eta(\textbf{r})=\frac{\rho_\alpha(\textbf{r})-\rho_\beta(\textbf{r})}{\rho_\text{C}(\textbf{r})}.
\end{align}
Substituting Eq. \eqref{spin_polari} into Eq. \eqref{pi_in_spin_polari} and rearranging the expression gives rise to an alternative expression for the spin polarization in terms of $\rho_\text{C}$ and $\pi$,
\begin{align}\label{pair_rho_s}
    \check{\rho}_\mathrm{S}(\textbf{r})=\rho_\text{C}(\textbf{r})\sqrt{1-\frac{4\pi(\textbf{r})}{\rho_\text{C}^2(\textbf{r})}}. 
\end{align}
This translation between the spin density, the total density, and the pair density has been made under the assumption that these quantities are constructed from a \emph{single} Slater determinant. 
A direct application utilizing a multiconfigurational wave function will, however, give rise to problems, as noted by \citeauthor{ziegler1977calculation}\cite{ziegler1977calculation} and later further examined by \citeauthor{becke1995extension}\cite{becke1995extension}. 
These problems manifest themselves for multiconfigurational wave functions as regions of space attaining a \emph{purely} imaginary contribution to the spin density.
 It is easily seen that this occurs if the fraction $\frac{4\pi(\textbf{r})}{\rho_\text{C}^2(\textbf{r})} $ 
 becomes larger than one.  It is an open question how this should be treated, both conceptually and practically. 
 It has been shown that this scenario in large-density regions is related to the ionic character of excited states\cite{hapka2020b},
 but in terms of Eq.~\eqref{pair_rho_s} it is  problematic in several ways, as will be illustrated.
Since the electron spin density is an observable, the spin density operator is Hermitian, meaning its eigenspectrum is included in the field of real numbers.
To circumvent this, \citeauthor{perdew1995escaping}\cite{perdew1995escaping} suggested that the spin density expressed from the pair density is rather an auxiliary quantity than an observable.
\citeauthor{becke1995extension}\cite{becke1995extension} argued that allowing $\check{\rho}_\mathrm{S}$ to have an imaginary component posed no problem for the exchange-correlation energy, since the exact functional only contains even powers of the spin density, thus resulting in a real energy contribution.
These arguments for $\check{\rho}_\mathrm{S}$ were utilized by \citeauthor{rodrigues2023multiconfigurational}\cite{rodrigues2023multiconfigurational} to realize the auxiliary spin density, thereby removing the discontinuity in Eq. \eqref{pair_rho_s}. 

\

In this work, we follow Li Manni \textit{et al.}\cite{li2014multiconfiguration} and  substitute the spin density occurrences within the usual short-range exchange-correlation functionals (Eq. \eqref{energy_density}) with the pair density dependent expression of the spin polarization, Eq. \eqref{pair_rho_s}: 
$\rho_\mathrm{S}\rightarrow \Re(\check{\rho}_\mathrm{S}$). 
The exchange-correlation energy can thus be expressed for a short-range \emph{translated} on-top pair local density approximation (sr$t$LDA) functional as,
\begin{align}\label{exc_pair_density}
    E^{\mathrm{sr}}_{\mathrm{xc}}\left[\rho_\text{C}(\textbf{r},\boldsymbol{\lambda}),\pi(\textbf{r},\boldsymbol{\lambda})\right] &=\int\text{e}^\mathrm{sr}_{\mathrm{xc}}\left(\rho_\text{C}(\textbf{r},\boldsymbol{\lambda}),\pi(\textbf{r},\boldsymbol{\lambda})\right)\,\text{d}\textbf{r}.
\end{align}
With this translation of the short-range functional, all new contributions needed to extend the MC-srDFT model into the MC-srPDFT model will emerge from the short-range exchange-correlation terms. 
The new additional expressions (not occuring in MC-srDFT) needed for a fully variational direct second-order optimization of the MC-srPDFT model will be the focus of the following Section.  

\section{Second-order optimization of the electronic MC-srPDFT energy}
To find an electronic stationary point (minimum if electronic ground state) on the energy hypersurface, we employ the restricted-step second-order optimization algorithm implemented in the \texttt{DALTON} program.\cite{jensen1986, jensen1987} 
To derive the explicit contributions, we expand the electronic energy in Eq.~\eqref{energy_ex_H_xc} to  second order in the wave function parameters as a Taylor series,
\begin{align}\label{taylor_series}
E(\boldsymbol{\lambda})=E_{0}+\boldsymbol{\lambda}^{\top}\boldsymbol{E}^{[1]}+\frac{1}{2}\boldsymbol{\lambda}^{\top}\boldsymbol{E}^{[2]}\boldsymbol{\lambda} + \cdots ,
\end{align}
where $E_{0}$ is the energy at the current set of wave function parameters.
We denote this point on the hypersurface as the current expansion point (CEP), i.e., $\boldsymbol{\lambda}^\top=(\boldsymbol{0}^\top, \boldsymbol{0}^\top)$. At this point, the energy is simply
\begin{align}
    E_{0}=E^\text{lr}(\boldsymbol{0}) + E_\text{H}^\text{sr}\left[\rho_\text{C}(\textbf{r},\boldsymbol{0})\right]+ E_\text{xc}^\text{sr}\left[\xi(\textbf{r},\boldsymbol{0})\right].
\end{align}
Note that $\textbf{c}=\boldsymbol{0}$ is simply achieved by updating the set of configuration coefficients in each optimization cycle, while $\boldsymbol{\kappa}=\boldsymbol{0}$ corresponds to doing an integral transformation each time the new set of parameters is incorporated such that the orbitals entering the molecular integrals correspond to the molecular orbitals for that specific CEP. 
The simultaneous optimization of the configuration coefficients and orbital rotation parameters thus becomes non-linear.
Note that $|0\rangle$ is the state comprised of the wave function parameters at the CEP, and this state is, therefore, not \emph{necessarily} the ground-state wave function. 
We are interested in determining the gradient, $\boldsymbol{E}^{[1]}$, and Hessian, $\boldsymbol{E}^{[2]}$, of the electronic energy to determine the optimal step. 
By inserting the MC-srDFT energy expression (Eq.~\ref{energy_ex_H_xc}) in Eq.~\eqref{taylor_series}, we identify the following energy contributions for the 0'th through 2'nd order,
\begin{subequations}
\begin{align}\label{second_order_ea}
    E &= E^\text{lr} + E_\text{H}^\text{sr}+ E_\text{xc}^\text{sr}\\
    &+ \sum_i\left(\frac{\partial E^\text{lr}}{\partial\lambda_i} + \frac{\partial E^\text{sr}_\text{H}}{\partial\lambda_i} + \frac{\partial E^\text{sr}_\text{xc}}{\partial\lambda_i}\right)\lambda_i\label{second_order_eb}\\
    &+ \frac{1}{2}\sum_i\sum_j\lambda_j\left(\frac{\partial^2 E^\text{lr}}{\partial\lambda_j\partial\lambda_i} + \frac{\partial^2 E^\text{sr}_\text{H}}{\partial\lambda_j\partial\lambda_i} + \frac{\partial^2 E^\text{sr}_\text{xc}}{\partial\lambda_j\partial\lambda_i}\right)\lambda_i + \cdots\label{second_order_ec} .
\end{align}
\end{subequations}
As a consequence of the introduction of the on-top pair density in the short-range exchange-correlation functional instead of the spin density, it is clear that new terms emerge in all orders (0'th through 2'nd) within the third term of Eqs. \eqref{second_order_ea}-\eqref{second_order_ec}. 
Whilst the charge part of the functional will remain unchanged, the spin-polarization part will change with the substitution of the spin density $\rho_\mathrm{S}(\textbf{r})$ with $\check{\rho}_\mathrm{S}(\textbf{r})$ in Eq.~\eqref{pair_rho_s}.
In the next Subsections, we turn our attention to the new exchange-correlation gradient and Hessian terms necessary for a second-order optimization scheme.
Discussion about the implementation of the long-range and short-range Hartree terms can be found in Ref.\ \onlinecite{hedegaard2018multiconfigurational}.
We give the expressions for sr$t$LDA, the local spin density approximation (srLSDA) translated using Eq.\ \eqref{pair_rho_s}.
The gradient and Hessian expressions below are the same for any other srPLDA model and can straightforwardly be extended to short-range GGA models based on the on-top pair density. A method that ensures  this extension leads to continueous first- and second-order derivatives can be found in Ref.~\citenum{carlson2015}.

\subsection{Exchange-correlation contributions to the gradient: the consequences of the on-top pair density}

Taking the first-order derivative of the on-top pair density-dependent short-range exchange-correlation energy functional with respect to a general wave function parameter gives rise to the following gradient expression,
\begin{align}
    \frac{\partial E_{\text{xc}}^{\text{sr}}[\rho_\text{C},\pi]}{\partial\lambda_{i}}
    &=\int\left(\frac{\partial\text{e}_{\text{xc}}^{\text{sr}}(\rho_\text{C}(\textbf{r},\boldsymbol{\lambda}),\pi(\textbf{r},\boldsymbol{\lambda}))}{\partial\rho_\text{C}(\textbf{r},\boldsymbol{\lambda})}\frac{\partial\rho_\text{C}(\textbf{r},\boldsymbol{\lambda})}{\partial\lambda_{i}}\right. \nonumber\\ & \quad+\left.\frac{\partial\text{e}_{\text{xc}}^{\text{sr}}(\rho_\text{C}(\textbf{r},\boldsymbol{\lambda}),\pi(\textbf{r},\boldsymbol{\lambda}))}{\partial\pi(\textbf{r},\boldsymbol{\lambda})}\frac{\partial\pi(\textbf{r},\boldsymbol{\lambda})}{\partial\lambda_{i}}\right)\,\text{d}\textbf{r} ,
\end{align}
where the new types of terms are the derivative of the energy density, $\text{e}^\text{sr}_\text{xc}$, with the pair density and the derivative of the pair density with respect to the wave function parameters.
By inserting the explicit expression for the on-top pair density (Eq.\eqref{pair_density_orbs}), we define the effective one- and two-electron singlet gradient operators as,
\begin{subequations}
\begin{align}
    ^{1\text{e}}\hat{V}^\mathrm{sr}_\text{xc} &=\sum_{pq}\left(\int\frac{\partial\text{e}_{\text{xc}}^{\text{sr}}(\rho_\text{C},\pi)}{\partial\rho_\text{C}}\Omega_{pq}\,\text{d}\textbf{r}\right)\hat{E}_{pq}\nonumber\\
    &= \sum_{pq} {}^{1\text{e}}V^\mathrm{sr}_{\text{xc},\{pq\}}\hat{E}_{pq} \label{grad_1el_op}\\
    ^{2\text{e}}\hat{V}^\mathrm{sr}_\text{xc} &=\sum_{pq,rs}\left(\int\frac{\partial\text{e}_{\text{xc}}^{\text{sr}}(\rho_\text{C},\pi)}{\partial\pi}\Omega_{pq}\Omega_{rs}\,\text{d}\textbf{r}\right)\hat{e}_{pq,rs}\nonumber\\
    &= \sum_{pq,rs} {}^{2\text{e}}V^\mathrm{sr}_{\text{xc},\{pq,rs\}}\hat{e}_{pq,rs} . \label{grad_2el_op}
\end{align}
\end{subequations}
The explicit expressions for the gradient with respect to the orbital rotation coefficients and the configuration coefficients can now be written as
\begin{align}\label{pair_orb_grad}
    E^{\mathrm{sr},[1]}_{\mathrm{xc},pq} = \frac{\partial E_{\text{xc}}^{\text{sr}}[\rho_\text{C},\pi]}{\partial\kappa_{pq}}\Bigg|_{\boldsymbol{\lambda=0}} & =\langle0|\left[\hat{E}_{pq}^-\,,\,^{1\text{e}}\hat{V}^\mathrm{sr}_\text{xc} + ^{2\text{e}}\hat{V}^\mathrm{sr}_\text{xc}\,\right]|0\rangle , 
\end{align}
and
\begin{align}\label{pair_ci_grad}
    E^{\mathrm{sr},[1]}_{\mathrm{xc},i} = \frac{\partial E_{\text{xc}}^{\text{sr}}[\rho_\text{C},\pi]}{\partial c_{i}}\Bigg|_{\boldsymbol{\lambda=0}} &= 2  \Bigl( \langle i|\,^{1\text{e}}\hat{V}^\mathrm{sr}_\text{xc} + ^{2\text{e}}\hat{V}^\mathrm{sr}_\text{xc}|0\rangle\nonumber\\
    &\quad-C_{i}^{(0)}\,\langle0|^{1\text{e}}\hat{V}^\mathrm{sr}_\text{xc} + ^{2\text{e}}\hat{V}^\mathrm{sr}_\text{xc} |0\rangle \Big).
\end{align}
The expressions with the effective operator $^{1\text{e}}\hat{V}^\mathrm{sr}_\text{xc}$ is almost identical (except for the dependence of on-top pair density in the functional derivative)  to the corresponding equations for the gradient in Ref.~\citenum{hedegaard2018multiconfigurational}. However, the term including $^{2\text{e}}\hat{V}^\mathrm{sr}_\text{xc}$ is new to the MC-srPDFT model and comes in as a consequence of the two-electron on-top pair density.

\subsection{Exchange-correlation contributions to the Hessian: the consequences of the on-top pair density}

To avoid the explicit construction of the full Hessian matrix, the implementation employs a direct iterative scheme utilizing so-called trial vectors. 
The full Hessian is thus projected onto a subspace, $\textbf{H}'$, spanned by these trial vectors, which significantly reduces the memory requirements. 
For $n$ trial vectors, the subspace Hessian can be written as
\begin{align}\label{subspace_hamiltonian}
\textbf{H}'=&
\left(\boldsymbol{\sigma}_1 \dots \boldsymbol{\sigma}_n  \right){}^\top 
\left( \textbf{b}_1 \dots \textbf{b}_n \right),
 \textrm{ i.e. }H'_{kl} = \boldsymbol{\sigma}_k^\top \textbf{b}_l ,
\end{align}
where the $\boldsymbol{\sigma}$ vectors contain the results of the direct matrix-vector product of the electronic Hessian, $\textbf{E}^{[2]}$, with the trial vectors $\textbf{b}$,
\begin{align}\label{sigma_vec}
\boldsymbol{\sigma}_i=\textbf{E}^{[2]} \textbf{b}_i = 
\left[\begin{array}{cc}
        \textbf{E}_{cc}^{[2]} & \textbf{E}_{co}^{[2]}\\
        \textbf{E}_{oc}^{[2]} & \textbf{E}_{oo}^{[2]}
        \end{array}\right]\left[\begin{array}{c}
        \textbf{b}^{c}_i\\
        \textbf{b}^{o}_i
    \end{array}\right] . 
\end{align}

The trial vectors are constructed as implemented in \texttt{DALTON}.\cite{jensen1986,jensen1987} 

Due to the multiconfigurational wave function being fully determined by the configuration coefficients and the occupied orbitals, the trial vectors consist of a configuration (\emph{c}) part and an orbital (\emph{o}) part. 
This blocking has been shown explicitly, because the four sub-blocks of the electronic Hessian require different algorithms.
The subscripts of the four sub-blocks of the electronic Hessian denote the variables of which the derivative of the electronic energy is taken. 
The full form is shown in the parenthesis of Eq.~\eqref{second_order_ec}. 
Here we focus on the new short-range exchange-correlation part which generally can be written as (we purposely omit noting the evaluation at $\boldsymbol{\lambda=0}$, however, evaluation at the CEP wave function parameters should be kept in mind)
\begin{align}
   \frac{\partial^{2}E_{\text{xc}}^{\text{sr}}[\xi]}{\partial\lambda_{i}\partial\lambda_{j}} & =\int\Bigg[\frac{\partial^{2}\text{e}_{\text{xc}}^{\text{sr}}(\rho_\mathrm{C},\pi)}{\partial\rho_\mathrm{C}^{2}}\frac{\partial\rho_\mathrm{C}}{\partial\lambda_{j}}\frac{\partial\rho_\mathrm{C}}{\partial\lambda_{i}}\nonumber\\
    &\quad+\frac{\partial\text{e}_{\text{xc}}^{\text{sr}}(\rho_\mathrm{C},\pi)}{\partial\rho_\mathrm{C}}\frac{\partial^{2}\rho_\mathrm{C}}{\partial\lambda_{i}\partial\lambda_{j}} \nonumber\\
    &\quad+\frac{\partial^{2}\text{e}_{\text{xc}}^{\text{sr}}(\rho_\mathrm{C},\pi)}{\partial\pi^{2}}\frac{\partial\pi}{\partial\lambda_{j}}\frac{\partial\pi}{\partial\lambda_{i}}\nonumber\\
    &\quad+\frac{\partial\text{e}_{\text{xc}}^{\text{sr}}(\rho_\mathrm{C},\pi)}{\partial\pi}\frac{\partial^{2}\pi}{\partial\lambda_{i}\partial\lambda_{j}} \nonumber\\
    &\quad+\frac{\partial^2\text{e}_{\text{xc}}^{\text{sr}}(\rho_\mathrm{C},\pi)}{\partial\rho_\mathrm{C}\partial\pi}\left(\frac{\partial\rho_\mathrm{C}}{\partial\lambda_{j}}\frac{\partial\pi}{\partial\lambda_{i}}+\frac{\partial\rho_\mathrm{C}}{\partial\lambda_{i}}\frac{\partial\pi}{\partial\lambda_{j}}\right)\Bigg]\,\text{d}\textbf{r}.
    \label{hessian_def}
\end{align}
Due to the non-linearity of the exchange-correlation functional, both linear and non-linear terms dependent on the charge and on-top pair densities emerge. 
By contraction with a general trial vector, the practical implementation can be simplified by constructing transformed density matrices, i.e., one-index transformed and transition density matrices for an orbital or configuration trial vector, respectively.
The explicit form of these intermediary density matrices will be investigated in a later subsection.
A contraction of the short-range electronic Hessian term in Eq. ~\eqref{hessian_def} with a general trial vector can be written as
\begin{align}\label{contraction_b}
    \sum_{j}\frac{\partial^{2}E_{\mathrm{xc}}^{\mathrm{sr}}\left[\xi\right]}{\partial\lambda_{i}\partial\lambda_{j}}b_{j}^{\lambda} = \int\Bigg(
    &\frac{\partial^{2}\mathrm{e}_{\mathrm{xc}}^{\mathrm{sr}}\left(\rho_{\mathrm{C}},\pi\right)} {\partial\rho_{\mathrm{C}}^{2}}  \Bigg[\sum_j \frac{\partial\rho_{\mathrm{C}}}{\partial\lambda_{j}}b^{\lambda}_j \Bigg] \frac{\partial\rho_{\mathrm{C}}}{\partial\lambda_{i}}  \nonumber\\
    + &\frac{\partial^{2}\mathrm{e}_{\mathrm{xc}}^{\mathrm{sr}}\left(\rho_{\mathrm{C}},\pi\right)}{\partial\rho_{\mathrm{C}}\partial\pi} \Bigg[\sum_j \frac{\partial\pi}{\partial\lambda_{j}}b^{\lambda}_j \Bigg] \frac{\partial\rho_{\mathrm{C}}}{\partial\lambda_{i}}  \nonumber\\
    +&\frac{\partial^{2}\mathrm{e}_{\mathrm{xc}}^{\mathrm{sr}}\left(\rho_{\mathrm{C}},\pi\right)}{\partial\pi^{2}}\Bigg[\sum_j \frac{\partial\pi}{\partial\lambda_{j}}b^{\lambda}_j \Bigg] \frac{\partial\pi}{\partial\lambda_{i}} \nonumber\\
    + &\frac{\partial^{2}\mathrm{e}_{\mathrm{xc}}^{\mathrm{sr}}\left(\rho_{\mathrm{C}},\pi\right)}{\partial\rho_{\mathrm{C}}\partial\pi} \Bigg[\sum_j \frac{\partial\rho_{\mathrm{C}}}{\partial\lambda_{j}}b^{\lambda}_j \Bigg]  \frac{\partial\pi}{\partial\lambda_{i}}    \nonumber\\
    +&\frac{\partial\mathrm{e}_{\mathrm{xc}}^{\mathrm{sr}}\left(\rho_{\mathrm{C}},\pi\right)}{\partial\rho_{\mathrm{C}}}\Bigg[\sum_{j}\frac{\partial^{2}\rho_{\mathrm{C}}}{\partial\lambda_{i}\partial\lambda_{j}}b_{j}^{\lambda}\Bigg] \nonumber\\
    + &\frac{\partial\mathrm{e}_{\mathrm{xc}}^{\mathrm{sr}}\left(\rho_{\mathrm{C}},\pi\right)}{\partial\pi}\Bigg[\sum_{j}\frac{\partial^{2}\pi}{\partial\lambda_{i}\partial\lambda_{j}}b_{j}^{\lambda}\Bigg]\Bigg) \,\mathrm{d}\textbf{r}.
\end{align}
The first four terms are the terms non-linear in $\rho_{\mathrm{C}}$ and $\pi$, which require special attention. The last two terms are linear in $\rho_{\mathrm{C}}$ and $\pi$ and can be calculated with the effective operators in Eqs.~\eqref{grad_1el_op}-\eqref{grad_2el_op} defined for the gradient evaluation, as shown below.
We can simplify the expressions for the non-linear terms in two steps: First, we define (for a general trial vector), the linearly transformed density gradients,  $\tilde{\rho}_{\mathrm{C}}^{\lambda}$, and $\tilde{\pi}^{\lambda}$ (where the superscript “$\lambda$” indicates if the transformation is of orbital or configuration type) for the charge density and on-top pair densities:
\begin{align}\label{linear-transformed-dens}
\tilde{\rho}_{\mathrm{C}}^{\lambda}\equiv\left[\sum_{j}\frac{\partial\rho_{\mathrm{C}}}{\partial\lambda_{j}}b_{j}^{\lambda}\right]_{\boldsymbol{\lambda}=\boldsymbol{0}}\, \, \, \text{and}\, \, \, \, \, \, 
\tilde{\pi}^{\lambda}\equiv\left[\sum_{j}\frac{\partial\pi}{\partial\lambda_{j}}b_{j}^{\lambda}\right]_{\boldsymbol{\lambda}=\boldsymbol{0}} .
\end{align}
Second, we utilize these linearly transformed charge and on-top pair densities to define effective one-electron and two-electron srPDFT operators as,
\begin{subequations}
    \begin{align}
        ^{1\mathrm{e}}\hat{V}_{\mathrm{xc}}^{\mathrm{sr},\lambda} 
        &=\sum_{pq} {}^{1\mathrm{e}}V_{\mathrm{xc},pq}^{\mathrm{sr},\lambda}\,\hat{E}_{pq}\label{hes_1el_op}\\
        ^{2\mathrm{e}}\hat{V}_{\mathrm{xc}}^{\mathrm{sr},\lambda}
        &=\sum_{pqrs} {}^{2\mathrm{e}}V_{\mathrm{xc},pqrs}^{\mathrm{sr},\lambda}\,\hat{e}_{pqrs}\label{hes_2el_op},
    \end{align}
\end{subequations}
with
\begin{subequations}
    \begin{align}
        ^{1\mathrm{e}}V_{\mathrm{xc},pq}^{\mathrm{sr},\lambda} 
        &=\int\left(\frac{\partial^{2}\mathrm{e}_{\mathrm{xc}}^{\mathrm{sr}}\left(\rho_{\mathrm{C}},\pi\right)}{\partial\rho_{\mathrm{C}}^{2}}\tilde{\rho}_{\mathrm{C}}^{\lambda}+\frac{\partial^{2}\mathrm{e}_{\mathrm{xc}}^{\mathrm{sr}}\left(\rho_{\mathrm{C}},\pi\right)}{\partial\rho_{\mathrm{C}}\partial\pi}\tilde{\pi}^{\lambda}\right)\Omega_{pq}\mathrm{d}\textbf{r} \label{hes_1el_op_integrals} \\
        ^{2\mathrm{e}}V_{\mathrm{xc},pqrs}^{\mathrm{sr},\lambda} 
        &=\int\left(\frac{\partial^{2}\mathrm{e}_{\mathrm{xc}}^{\mathrm{sr}}\left(\rho_{\mathrm{C}},\pi\right)}{\partial\pi^{2}}\tilde{\pi}^{\lambda}+\frac{\partial^{2}\mathrm{e}_{\mathrm{xc}}^{\mathrm{sr}}\left(\rho_{\mathrm{C}},\pi\right)}{\partial\rho_{\mathrm{C}}\partial\pi}\tilde{\rho}_{\mathrm{C}}^{\lambda}\right)\nonumber\\ \quad&\times\Omega_{pq}\Omega_{rs}\mathrm{d}\textbf{r} . \label{hes_2el_op_integrals}
    \end{align}
\end{subequations}    
Utilizing the definitions in Eqs. \eqref{hes_1el_op}-\eqref{hes_2el_op_integrals} and the definitions of $\rho_{\text{C}}$ and $\pi$ in Eqs.\eqref{density-exp-value} and \eqref{pi_sq}, respectively, we obtain:
\begin{align}
 &\sum_{j}\frac{\partial^{2}E_{\mathrm{xc}}^{\mathrm{sr}}\left[\xi\right]}{\partial\lambda_{i}\partial\lambda_{j}}b_{j}^{\lambda} =  \frac{\partial \langle \Psi^\text{lr}(\boldsymbol{\lambda})| ^{1\mathrm{e}}\hat{V}_{\mathrm{xc}}^{\mathrm{sr},\lambda} + ^{2\mathrm{e}}\hat{V}_{\mathrm{xc}}^{\mathrm{sr},\lambda}  |\Psi^\text{lr}(\boldsymbol{\lambda})\rangle }{\partial\lambda_{i}}  \nonumber\\
    &\quad+\Bigg[\sum_{j}\frac{\partial^{2} \langle \Psi^\text{lr}(\boldsymbol{\lambda})| ^{1\mathrm{e}}\hat{V}_{\mathrm{xc}}^{\mathrm{sr}} + ^{2\mathrm{e}}\hat{V}_{\mathrm{xc}}^{\mathrm{sr}}  |\Psi^\text{lr}(\boldsymbol{\lambda})\rangle } {\partial\lambda_{i}\partial\lambda_{j}}b_{j}^{\lambda}\Bigg], 
\end{align}
where we have also used the two effective operators defined in Eqs.~\eqref{grad_1el_op} and \eqref{grad_2el_op}. The structure of the Hessian is similar to the MC-srDFT method (see e.g.~Ref. \citenum{hedegaard2018multiconfigurational}), where the first term is gradient-like. 
This gradient-like structure is unique to models with multiconfigurational methods that involve a non-linear potential and it also occurs in previous multiconfigurational DFT hybrids.
The second term has the usual structure known from the pure MCSCF electronic Hessian. 
The effective one-electron operators $^{1\mathrm{e}}\hat{V}_{\mathrm{xc}}^{\mathrm{sr}}$ and $^{1\mathrm{e}}\hat{V}_{\mathrm{xc}}^{\mathrm{sr},\lambda}$ are related to the effective operators in the Hessian for MC-srDFT (except for the kernel involving a $\pi$ variable as well as the inclusion of a transformed on-top pair density). 
The new contribution types to the Hessian is a direct consequence of the two-electron on-top pair density and enters in the form of the effective two-electron operators  $^{2\mathrm{e}}\hat{V}_{\mathrm{xc}}^{\mathrm{sr}} $ and $^{2\mathrm{e}}\hat{V}_{\mathrm{xc}}^{\mathrm{sr},\lambda}$. 
These operators have a unique structure that does not appear in the previous MC-srDFT models. 

\ 

\textbf{Linearly transformed densities.} 
The explicit form of the linearly transformed densities (Eq.~\ref{linear-transformed-dens}) used to construct the integrals in  Eqs.~\eqref{hes_1el_op_integrals} and \eqref{hes_2el_op_integrals} will depend on whether we use a configuration or an orbital trial vector. 
For a configuration trial vector, expressions for the linearly transformed charge density gradient can be explicitly derived as
\begin{align}
 \tilde{\rho}^{c}_{\mathrm{C}}   = \left[\sum_{j}\frac{\partial\rho_{\mathrm{C}}}{\partial c_{j}}b_{j}^{c}\right]_{\boldsymbol{\lambda}=\boldsymbol{0}}
    &=2\sum_{pq}\Omega_{pq}\langle 0|\hat{E}_{pq}|B\rangle ,
\end{align}
with $|B\rangle = \sum_j | j \rangle b^c_j$.
An expression for the transformed gradient on-top pair density can similarly be derived as
\begin{align}
     \tilde{\pi}^{c} = \left[\sum_{j}\frac{\partial\pi}{\partial c_{j}}b_{j}^{c}\right]_{\boldsymbol{\lambda}=\boldsymbol{0}}
    &=2\sum_{pqrs}\Omega_{pq}\Omega_{rs}\langle 0|\hat{e}_{pq,rs}|B\rangle . 
\end{align} 
The state $|B\rangle$ is by construction orthogonal to the MC state $|0\rangle$, and thus non-redundant, in agreement with the projector $\mathcal{P}$ in Eq.\ \eqref{wavefun_ansatz}. 
Similarly, for an orbital trial vector the contributions from the charge and on-top pair densities to the orbital gradient can  be determined as
\begin{align}
     \tilde{\rho}^{o}_{\mathrm{C}} = \left[\sum_{t>u}\frac{\partial\rho_{\mathrm{C}}}{\partial\kappa_{tu}}b_{tu}^{o}\right]_{\boldsymbol{\lambda}=\boldsymbol{0}} 
    &=\sum_{pq}\left(\Omega_{\tilde{p}q}+\Omega_{p\tilde{q}}\right)\langle0|\hat{E}_{pq}|0\rangle\nonumber\\
    &=2\sum_{pq}\Omega_{\tilde{p}q}\langle0|\hat{E}_{pq}|0\rangle, \label{eq:charge_orb_gradient}
\end{align}
and 
\begin{align}
         \tilde{\pi}^{o} =\left[\sum_{t>u}\frac{\partial\pi}{\partial\kappa_{tu}}b_{tu}^{o}\right]_{\boldsymbol{\lambda}=\boldsymbol{0}}&=\sum_{pqrs}\left(\Omega_{\tilde{p}q,rs}+\Omega_{p\tilde{q},rs}+\Omega_{pq,\tilde{r}s}+\Omega_{pq,r\tilde{s}}\right)\nonumber\\
    &\quad\times\langle0|\hat{e}_{pqrs}|0\rangle\nonumber\\
    &=2\sum_{pqrs}\left(\Omega_{\tilde{p}q,rs}+\Omega_{p\tilde{q},rs}\right)\langle0|\hat{e}_{pqrs}|0\rangle,\label{eq:spin_orb_gradient}
\end{align} 
where we have defined a short-hand notation for the one-index transformed orbital overlap distributions 
\begin{align}
\Omega_{\tilde{p}q} &=\sum_{t}\left[b_{pt}^{o}\phi_{t}\right]\phi_{q}, \label{1ind_omegapq} \\
\Omega_{\tilde{p}q,rs} &=\sum_{t}\left[b_{pt}^{o}\phi_{t}\right]\phi_{q}\phi_{r}\phi_{s}.\label{1ind_omegapqrs}
\end{align}
Note that the expression for $\tilde{\pi}^{o}$ cannot be further reduced because in general $\langle0|\hat{e}_{pqrs}|0\rangle\neq \langle0|\hat{e}_{qprs}|0\rangle$.

\ 

\textbf{Sigma vector contributions.} Utilizing the definitions of the above srPDFT gradient and Hessian operators (Eqs. \eqref{grad_1el_op}-\eqref{grad_2el_op} and Eqs.\eqref{hes_1el_op}-\eqref{hes_2el_op}, respectively) the short-range \emph{xc} terms contributing to the four sigma vectors of Eq. \eqref{sigma_vec} can now be written as
\begin{subequations}
    \begin{align}
        \sigma_{i}^{\text{sr},cc} &= 2\left(\langle i|^{1\mathrm{e}}\hat{V}^{\mathrm{sr},c}_{\mathrm{xc}}+{}^{2\mathrm{e}}\hat{V}_{\mathrm{xc}}^{\mathrm{sr},c}|0\rangle-C_{i}^{(0)}\,\langle0|^{1\mathrm{e}}\hat{V}_{\mathrm{xc}}^{\mathrm{sr},c}+{}^{2\mathrm{e}}\hat{V}_{\mathrm{xc}}^{\mathrm{sr},c}|0\rangle\right)\nonumber\\
        &\quad+2\left(\langle i|^{1\mathrm{e}}\hat{V}^\mathrm{sr}_{\mathrm{xc}}+{}^{2\mathrm{e}}\hat{V}^\mathrm{sr}_{\mathrm{xc}}|B\rangle-b_i^c\,\langle0|^{1\mathrm{e}}\hat{V}^\mathrm{sr}_{\mathrm{xc}}+{}^{2\mathrm{e}}\hat{V}^\mathrm{sr}_{\mathrm{xc}}|0\rangle\right)\label{sigma_cc}\\
        \sigma_{i}^{\text{sr},co} & =2\langle i|^{1\mathrm{e}}\hat{V}_{\mathrm{xc}}^{\mathrm{sr},o}+{}^{2\mathrm{e}}\hat{V}_{\mathrm{xc}}^{\mathrm{sr},o}+{}^{1\mathrm{e}}\hat{\tilde{V}}^\mathrm{sr}_{\mathrm{xc}}+{}^{2\mathrm{e}}\hat{\tilde{V}}^\mathrm{sr}_{\mathrm{xc}}|0\rangle\nonumber \\
        &\quad-2C_{i}^{(0)}\,\langle0|^{1\mathrm{e}}\hat{V}_{\mathrm{xc}}^{\mathrm{sr},o}+{}^{2\mathrm{e}}\hat{V}_{\mathrm{xc}}^{\mathrm{sr},o}+{}^{1\mathrm{e}}\hat{\tilde{V}}^\mathrm{sr}_{\mathrm{xc}}+{}^{2\mathrm{e}}\hat{\tilde{V}}^\mathrm{sr}_{\mathrm{xc}}|0\rangle\label{sigma_co}\\
        \sigma_{pq}^{\text{sr},oc} &=\langle0|\left[\hat{E}_{pq}^{-} \,,\,^{1\mathrm{e}}\hat{V}_{\mathrm{xc}}^{\mathrm{sr},c}+{}^{2\mathrm{e}}\hat{V}^{\mathrm{sr},c}_{\mathrm{xc}}\right]|0\rangle\nonumber\\
        &\quad+2\langle B|\left[\hat{E}_{pq}^{-}\,,\, ^{1\mathrm{e}}\hat{V}^\mathrm{sr}_{\mathrm{xc}}+{}^{2\mathrm{e}}\hat{V}^\mathrm{sr}_{\mathrm{xc}}\right]|0\rangle \label{sigma_oc}\\
        \sigma_{pq}^{\text{sr},oo} &=\langle0|\left[\hat{E}_{pq}^{-}\,,\,^{1\mathrm{e}}\hat{\tilde{V}}^\mathrm{sr}_{\mathrm{xc}}+{}^{2\mathrm{e}}\hat{\tilde{V}}^\mathrm{sr}_{\mathrm{xc}}+{}^{1\mathrm{e}}\hat{V}_{\mathrm{xc}}^{\mathrm{sr},o}+{}^{2\mathrm{e}}\hat{V}_{\mathrm{xc}}^{\mathrm{sr},o}\right]|0\rangle \nonumber\\
        &\quad+\frac{1}{2}\sum_t\left(E^{\mathrm{sr},[1]}_{\mathrm{xc},tq}\,b^o_{pt}-E^{\mathrm{sr},[1]}_{\mathrm{xc},pt}\,b^o_{qt}\right)\label{sigma_oo} , 
    \end{align}
\end{subequations}
where $E^{\mathrm{sr}[1]}_{\mathrm{xc},pq}$ can be found in Eq.~\eqref{pair_orb_grad}.
In the above expressions, one-index transformations of the one-electron and two-electron gradient operators are utilized.
For the one-electron term this is expressed in terms of the effective operator
\begin{align}
    ^{1\mathrm{e}}\hat{\tilde{V}}^\mathrm{sr}_{\mathrm{xc}} 
    &=2\sum_{pq}\, ^{1\mathrm{e}}V^\mathrm{sr}_{\mathrm{xc},\{\tilde{p}q\}}\hat{E}_{pq},
\end{align}
where the one-index transformation has been performed on the MO amplitudes as
\begin{align}
     ^{1\mathrm{e}}V^\mathrm{sr}_{\mathrm{xc},\{\tilde{p}q\}}=\int\frac{\partial\mathrm{e}_{\mathrm{xc}}^{\mathrm{sr}}\left(\rho_{\mathrm{C}}\left(\boldsymbol{r},\boldsymbol{\lambda}\right),\pi\left(\boldsymbol{r},\boldsymbol{\lambda}\right)\right)}{\partial\rho_{\mathrm{C}}\left(\boldsymbol{r},\boldsymbol{\lambda}\right)}\Omega_{\tilde{p}q}\ \mathrm{d}\textbf{r}.
\end{align}
Similarly for the two-electron gradient,
\begin{align}
    ^{2\mathrm{e}}\hat{\tilde{V}}^\mathrm{sr}_{\mathrm{xc}}
    &=2\sum_{pqrs}\left( ^{2\mathrm{e}}V^\mathrm{sr}_{\mathrm{xc},\{\tilde{p}q,rs\}}+ ^{2\mathrm{e}}V^\mathrm{sr}_{\mathrm{xc},\{p\tilde{q},rs\}}\right)\hat{e}_{pqrs},
\end{align}
where the one-index transformed two-electron MO amplitudes are denoted as
\begin{align}
    ^{2\mathrm{e}}V^\mathrm{sr}_{\mathrm{xc},\{\tilde{p}q,rs\}}&=\int\frac{\partial\mathrm{e}_{\mathrm{xc}}^{\mathrm{sr}}\left(\rho_{\mathrm{C}}\left(\boldsymbol{r},\boldsymbol{\lambda}\right),\pi\left(\boldsymbol{r},\boldsymbol{\lambda}\right)\right)}{\partial\pi\left(\boldsymbol{r},\boldsymbol{\lambda}\right)}\Omega_{\tilde{p}q}\Omega_{rs}\,\mathrm{d}\textbf{r}.
\end{align}

\subsection{Implementation}
In this subsection we discuss the practical implementation of the new electronic gradient and Hessian terms, which arise from the short-range exchange-correlation contribution due to the explicit functional dependency on the on-top pair density.
Hitherto, only non-specific, general orbital indices have been used. 
Hereafter, we will adopt the orbital index convention from Ref. \citenum{jensen1994electron}, which denotes orbitals of the different molecular spin-orbital subspaces as follows:
\begin{align*}
    &p,q,r,s\quad &&\text{for \emph{general} orbitals}\\
    &i,j,k,l\quad &&\text{for \emph{inactive} orbitals}\\
    &u,v,x,y\quad &&\text{for \emph{active} orbitals}\\
    &a,b,c,d\quad &&\text{for \emph{secondary} orbitals}
\end{align*}
The practical implementation of both the electronic gradient and the electronic Hessian terms relies on the construction of short-range exchange-correlation Fock matrix contributions.
The total generalized Fock matrix within the MC-srPDFT model consists of the sum of long-range and short-range contributions.
In the practical implementation, the integration of the short-range exchange-correlation (Eq. \eqref{exc_pair_density}) is performed over a numerical grid
\begin{align}\label{exc_numerical}
    E^{\mathrm{sr}}_{\mathrm{xc}}\left[\rho_\text{C}(\textbf{r},\boldsymbol{\lambda}),\pi(\textbf{r},\boldsymbol{\lambda})\right] &=\int\text{e}^\mathrm{sr}_{\mathrm{xc}}\left(\rho_\text{C}(\textbf{r},\boldsymbol{\lambda}),\pi(\textbf{r},\boldsymbol{\lambda})\right)\,\text{d}\textbf{r}\nonumber\\
    &\approx \sum_g w_g\,\text{e}^\mathrm{sr}_{\mathrm{xc}}\left(\rho_\text{C}(\textbf{r}_g,\boldsymbol{\lambda}),\pi(\textbf{r}_g,\boldsymbol{\lambda})\right),
\end{align}
where the sum runs over all grid points $\textbf{r}_g$ with associated weight $w_g$. 
We can now express an element of the total generalized Fock matrix \cite{siegbahn1981complete} with the srDFT modifications as
\begin{align}\label{generalized_fock}
    F_{pq} &= \sum_r h_{qr}D_{pr}+2\sum_{rst}P_{pr,st}\left(qr|st\right)^\mathrm{lr} +F_{\mathrm{H},pq}^\mathrm{sr}+ F_{\mathrm{xc},pq}^\mathrm{sr}\nonumber\\
    &=F^\mathrm{lr}_{pq}+F^\mathrm{sr}_{\mathrm{H},pq}+\sum_g w_g\,f^\mathrm{sr}_{\mathrm{xc},\{pq,g\}}.
\end{align}
With the orbital subspace division in the CAS-srPDFT model, the generalized Fock matrix can be expressed in terms of auxiliary Fock matrices.
Expressions for the long-range auxiliary Fock matrices can be found in the work by \citeauthor{siegbahn1981complete}\cite{siegbahn1981complete}.
We define the short-range exchange-correlation auxiliary \emph{inactive} (I), \emph{active} (A), and \emph{Q} Fock matrices as follows:
\begin{subequations}
    \begin{align}
        ^\text{I}F^\mathrm{sr}_{\mathrm{xc},pq} &= \sum_gw_g\left[f^\mathrm{sr}_{\{pq\},g} + \sum_j\left(2f^\mathrm{sr}_{\{jj,pq\},g}-f^\mathrm{sr}_{\{jq,pj\},g}\right)\right]\nonumber\\
        &=\sum_gw_g\left[f^\mathrm{sr}_{\{pq\},g} + \sum_jf^\mathrm{sr}_{\{jj,pq\},g}\right]\label{inac_fock}\\
        ^\text{A}F^\mathrm{sr}_{\mathrm{xc},pq} &= \sum_gw_g\left[\sum_{uv}D_{uv}\left(f^\mathrm{sr}_{\{uv,pq\},g}-\frac{1}{2}f^\mathrm{sr}_{\{uq,pv\},g}\right)\right]\nonumber\\
        &=\sum_gw_g\left[\frac{1}{2}\sum_{uv}D_{uv}f^\mathrm{sr}_{\{uv,pq\},g}\right]\label{ac_fock}\\
        ^\text{Q}F^\mathrm{sr}_{\mathrm{xc},uq} &= \sum_gw_g\left[\sum_{vxy} P_{uv,xy}f^\mathrm{sr}_{\{qv,xy\},g}\right]\label{q_fock} .
    \end{align}
\end{subequations}
The above expressions are general in the sense that the terms $f^\mathrm{sr}$ are placeholders for all previous srPDFT operator integrals evaluated at each grid point. 
To elaborate on this, we explicitly write the full short-range exchange-correlation contributions to the Fock matrices needed for the electronic orbital gradient, as well as the new unique pair-density contributions to the two-electron part of the CI gradient, in the following sections.
Furthermore, we elaborate on the unique non-linear contributions to the Hessian as a consequence of the srPDFT potential.
Thereby substituting $f^\mathrm{sr}$ with the relevant srPDFT integrals previously derived.

\textbf{Orbital gradient}. 
With the orbital gradient expressed in terms of generalized Fock matrices
\begin{align}
    E^{[1]}_{rs}=2(F_{sr}-F_{rs}),
\end{align}
the explicit short-range exchange-correlation working equations can be obtained by substituting the one-electron ($f^\mathrm{sr}_{\{pq\},g}$) and two-electron terms ($f^\mathrm{sr}_{\{pq,rs\},g}$) with the sr-gradient operator integral of Eq. \eqref{grad_1el_op} and Eq. \eqref{grad_2el_op}, respectively.
\begin{align}
    f^\mathrm{sr}_{\{pq\},g} &\rightarrow  {}^{1\mathrm{e}}V_{\text{xc},\{pq\},g}\nonumber\\
    &=\left(\frac{\partial\text{e}_{\text{xc}}^{\text{sr}}(\rho_\text{C},\pi)}{\partial\rho_\text{C}}\Omega_{pq}\,\right)_g\label{1el_working_grad}\\
    f^\mathrm{sr}_{\{pq,rs\},g} &\rightarrow  {}^{2\mathrm{e}}V_{\text{xc},\{pq,rs\},g}\nonumber\\
    &=\left(\frac{\partial\text{e}_{\text{xc}}^{\text{sr}}(\rho_\text{C},\pi)}{\partial\pi}\Omega_{pq}\Omega_{rs}\,\right)_g . 
    \label{2el_working_grad}
\end{align}
The working equations for the generalized Fock matrix of a CAS wave function can now be expressed using auxilliary short-range matrixes and the Fock matrix expression by \citeauthor{siegbahn1981complete}\cite{siegbahn1981complete} together with the above substitutions
\begin{align}
    F^\mathrm{sr}_{\mathrm{xc},iq} &= 2\left(^\text{I}F^\mathrm{sr}_{\mathrm{xc},iq} + ^\text{A}F^\mathrm{sr}_{\mathrm{xc},iq}\right)\nonumber \\
    &= 2\sum_gw_g\Bigg( {}^{1\mathrm{e}}V^\mathrm{sr}_{\text{xc},\{iq\},g} + \sum_j {}^{2\mathrm{e}}V^\mathrm{sr}_{\text{xc},\{jj,iq\},g} \label{fock_iq}\nonumber\\
    &\quad+\frac{1}{2}\sum_{uv}D_{uv} {}^{2\mathrm{e}}V^\mathrm{sr}_{\text{xc},\{uv,iq\},g}\Bigg)\\ 
    F^\mathrm{sr}_{\mathrm{xc},uq} &=\,^\text{I}F^\mathrm{sr}_{\mathrm{xc},wq}\,D_{uw}+2\,^\text{Q}F^\mathrm{sr}_{\mathrm{xc},uq}\nonumber\\ 
    &=\sum_gw_g\Bigg(\sum_wD_{uw}\left[ {}^{1\mathrm{e}}V^\mathrm{sr}_{\{wq\},g} + \sum_j {}^{2\mathrm{e}}V^\mathrm{sr}_{\{jj,wq\},g}\right]\nonumber\\
    &\quad+2\sum_{vxy} P_{uv,xy} {\,}^{2\mathrm{e}}V^\mathrm{sr}_{\{qv,xy\},g}\Bigg)\label{fock_uq}.
\end{align}

\textbf{CI gradient}. For the CI gradient, working equations for the contributions of the srDFT potential operators can similarly be found by considering the expression for a regular CI gradient in MCSCF
\begin{align}\label{working_ci_grad}
    E^{[1]}_i = \sum_{pq}h_{pq}\langle 0|\hat{E}_{pq}|i\rangle + \frac{1}{2}\sum_{pqrs}g_{pqrs}\langle0|\hat{e}_{pqrs}|i\rangle .
\end{align}
Examining the integrals within Eq. \eqref{working_ci_grad}, the CAS partition of the orbital subspace reduces the above equation to the following 
\begin{align}
    E^{[1]}_i &= 2\sum_{uv}\phantom{}^\mathrm{I}F_{uv}\langle 0|\hat{E}_{uv}|i\rangle + \sum_{uvxy}g_{uvxy}\langle0|\hat{e}_{uvxy}|i\rangle\nonumber\\
    &\quad- 2C_i^{(0)}(E^{(0)}-\phantom{}^\mathrm{I}E), 
\end{align}
with $\phantom{}^\mathrm{I}E=\sum_i(h_{ii}+\phantom{}^\mathrm{I}F_{ii})$.
For completeness we write the total gradient expression with the explicit short-range functional contributions,
\begin{align}
    E^{[1]}_i &= 2\sum_{uv}(\phantom{}^\mathrm{I}F^\mathrm{lr}_{uv}+j_{uv}^{\mathrm{sr}} +\phantom{}^\mathrm{I}F^\mathrm{sr}_{uv})\langle 0|\hat{E}_{uv}|i\rangle\nonumber\\
    &\quad+ \sum_{uvxy}(g^\mathrm{lr}_{uvxy}+\sum_g w_g\, g^\mathrm{srP}_{\{uvxy\},g})\,\langle0|\hat{e}_{uvxy}|i\rangle\nonumber\\
    &\quad- 2C_i^{(0)}(E^{(0)}-\sum_i(h_{ii}+\phantom{}^\mathrm{I}F^\mathrm{lr}_{ii}+j_{ii}^\mathrm{sr}+\phantom{}^\mathrm{I}F^\mathrm{sr}_{ii})) .
    \label{cigrad_working_eq}
\end{align}
From the above expression it is clear that, besides contributions in the form of short-range auxiliary inactive Fock matrix contributions (as in the orbital gradient, Eq. \eqref{inac_fock}), new two-electron integral contributions to $g_{uvxy}$ occur due to the on-top pair density. 
The working equations for the new unique two-electron contributions to the CI gradient can be written by evaluating the two-electron operator integral (Eq. \eqref{grad_2el_op}) over active indices only at each grid point
\begin{align}
    g_{\{uvxy\},g}^\mathrm{srP} &= \left(\frac{\partial\text{e}_{\text{xc}}^{\text{sr}}(\rho_\text{C},\pi)}{\partial\pi}\Omega_{uv}\Omega_{xy}\right)_{g}\label{ci_fock}.
\end{align}

\textbf{Hessian}. The direct Hessian contributions in Eqs.~\eqref{sigma_cc}-\eqref{sigma_oo} are similarly determined from modified Fock matrix contributions.
As seen from Eq. \eqref{contraction_b}, the second-order derivative of the functional with respect to the wave function parameters leads to both linear and non-linear contributions due to the non-linear functional dependency on the densities.
Given the high level of similarity between the linear Hessian terms and the usual second-order MCSCF contributions to the sigma vectors, the focus will be on the working equations of the non-linear terms.
For the usual linear second-order MCSCF contributions we refer to Ref. \citenum{jensen1991sirius} (Appendix 8A-B). 

For both types of trial vectors, we compute the one-electron and two-electron operators (Eq. \eqref{hes_1el_op} and Eq. \eqref{hes_2el_op}) with integral elements (Eq. \eqref{hes_1el_op_integrals} and \eqref{hes_2el_op_integrals}) to determine the non-linear (\emph{nl}) sigma vector contributions.

The influence of the type of trial vectors on the non-linear contributions lies in the construction of the transformed density (Eq. \eqref{eq:charge_orb_gradient}) and the transformed on-top pair density (Eq. \eqref{eq:spin_orb_gradient}).
In the practical implementation, both the regular on-top pair density (Eq. \eqref{pair_density_orbs}) and the transformed on-top pair density (Eq. \eqref{linear-transformed-dens}) are constructed in the basis of occupied molecular orbitals.
Construction in AO basis similar to typical implementations of Kohn-Sham DFT would be computationally very expensive and memory intensive because it would have a total dimension $N_{AO}^4$, the same as the two-electron integrals.
For an orbital trial vector ($\lambda=o$), the transformed density at each grid-point is determined by performing a one-index transformation on the MO overlap distributions as follows:
\begin{subequations}
    \begin{align}
        \tilde{\rho}_{\mathrm{C},g}^o &= \sum_{pq}\Omega_{\tilde{p}q,g} D_{pq}\\
        \tilde{\pi}^o_g &=2\sum_{pqrs}(\Omega_{\tilde{p}qrs}+\Omega_{p\tilde{q}rs})_gP_{pqrs} .
    \end{align}
\end{subequations}
The explicit construction of the one-index transformed orbital distributions are simple matrix-vector multiplications, cf. Eqs.~\eqref{1ind_omegapq}-\eqref{1ind_omegapqrs}.
For a configuration trial vector ($\lambda=c$), the densities are constructed using the one-electron and two-electron transition density matrices, respectively:
\begin{subequations}
    \begin{align}
        \tilde{\rho}_{\mathrm{C},g}^c &= 2\sum_{pq}\Omega_{pq,g} \langle0|\hat{E}_{pq}|B\rangle=2\sum_{pq}\Omega_{pq,g} D^\mathrm{T}_{pq}\\
        \tilde{\pi}^c_g &=2\sum_{pqrs}\Omega_{pqrs,g}\langle0|\hat{e}_{pqrs}|B\rangle=2\sum_{pqrs}\Omega_{pqrs,g}P^\mathrm{T}_{pqrs} .
    \end{align}
\end{subequations}
Since these non-linear srDFT operators are used both in modified CI gradient terms (Eq. \eqref{sigma_cc} and  Eq. \eqref{sigma_co}, respectively) and in modified orbital gradient terms (Eq. \eqref{sigma_oc} and Eq. \eqref{sigma_oo}, respectively), the modified densities will contribute to the short-range Fock matrices,
\begin{subequations}
    \begin{align}
        f^{\mathrm{sr},nl}_{\{pq\},g} &\rightarrow {}^{1\mathrm{e}}V_{\mathrm{xc},\{pq\},g}^{\mathrm{sr},\lambda} \nonumber\\
        &=\sum_{pq}\left(\left(\frac{\partial^{2}\mathrm{e}_{\mathrm{xc}}^{\mathrm{sr}}\left(\rho_{\mathrm{C}},\pi\right)}{\partial\rho_{\mathrm{C}}^{2}}\tilde{\rho}_{\mathrm{C}}^\lambda+\frac{\partial^{2}\mathrm{e}_{\mathrm{xc}}^{\mathrm{sr}}\left(\rho_{\mathrm{C}},\pi\right)}{\partial\rho_{\mathrm{C}}\partial\pi}\tilde{\pi}^{\lambda}\right)\Omega_{pq}\right)_g\label{1el_nl_orbhes}\\
        f^{\mathrm{sr},nl}_{\{pq,rs\},g} &\rightarrow  {}^{2\mathrm{e}}\hat{V}_{\mathrm{xc},\{pqrs\},g}^{\mathrm{sr},\lambda}\nonumber\\
        &=\sum_{pqrs}\bigg(\left(\frac{\partial^{2}\mathrm{e}_{\mathrm{xc}}^{\mathrm{sr}}\left(\rho_{\mathrm{C}},\pi\right)}{\partial\pi^{2}}\tilde{\pi}^{\lambda}+\frac{\partial^{2}\mathrm{e}_{\mathrm{xc}}^{\mathrm{sr}}\left(\rho_{\mathrm{C}},\pi\right)}{\partial\rho_{\mathrm{C}}\partial\pi}\tilde{\rho}_{\mathrm{C}}^{\lambda}\right)\nonumber\\ \quad&\times\Omega_{pq}\Omega_{rs}\bigg)_g\label{2el_nl_orbhes} ,
    \end{align}
\end{subequations}
as well as the two-electron integrals,
\begin{align}
    g_{\{uvxy\},g}^{\mathrm{srP},nl} &= \bigg(\left(\frac{\partial^{2}\mathrm{e}_{\mathrm{xc}}^{\mathrm{sr}}\left(\rho_{\mathrm{C}},\pi\right)}{\partial\pi^{2}}\tilde{\pi}^{\lambda}+\frac{\partial^{2}\mathrm{e}_{\mathrm{xc}}^{\mathrm{sr}}\left(\rho_{\mathrm{C}},\pi\right)}{\partial\rho_{\mathrm{C}}\partial\pi}\tilde{\rho}_{\mathrm{C}}^{\lambda}\right)\nonumber\\ \quad&\times\Omega_{uv}\Omega_{xy}\bigg)_g\label{2el_nl_orbhes_ci}.
\end{align}
Substituting Eqs. \eqref{1el_nl_orbhes} through \eqref{2el_nl_orbhes} into the auxiliary short-range Fock matrices (Eqs. \eqref{inac_fock} through \eqref{q_fock}), these quantities are used to determine the total short-range Fock matrix contributions. This is achieved by further substituting these auxiliary Fock matrices into Eq. \eqref{fock_iq} and Eq. \eqref{fock_uq}, thus generating the non-linear contributions to Eq. \eqref{sigma_oc} and Eq. \eqref{sigma_oo}.
The non-linear contributions to Eq. \eqref{sigma_cc} and Eq. \eqref{sigma_co} can similarly be determined by substituting Eq. \eqref{2el_nl_orbhes_ci} into Eq. \eqref{cigrad_working_eq} and constructing the inactive short-range Fock matrix using Eqs. \eqref{1el_nl_orbhes} through \eqref{2el_nl_orbhes}, as described for the pure orbital part.

\subsection{Implementation of the sr\emph{\textbf{t}}LDA functional}

The implementation of the translated sr$t$LDA functional was based on the srLDA functional\cite{paziani2006local} with the correlation part being PW92\cite{Perdew1992-dj,Perdew2018-kt}.
The pair-density was introduced into the functional using the translation $\rho_\mathrm{S}(\textbf{r}_g)\rightarrow \Re(\check{\rho}_\mathrm{S}(\textbf{r}_g))$ of Eq. (\ref{pair_rho_s}).
The first and second derivative of the exchange-correlation kernel needed for the construction of the gradient and Hessian was derived by implementing the functional into SymPy\cite{sympy}, and using SymPy's functionality to get analytical expressions for the derivatives.
The mathematical expressions was converted to Fortran code using SymPy's \textit{fcode} feature.
As noted in Section \ref{Theory_ontop_pair}, 
the square root in Eq. \eqref{pair_rho_s} can become imaginary for multiconfigurational wave functions,
also for regions of space with significant electronic density for certain states with ionic character.\cite{hapka2020b}
For the systems studied in this paper, imaginary values only appeared in regions of insignificant electron density.

\section{Computational Details}\label{ComDetails}
The presented equations of the previous sections for MC-sr$t$LDA have been implemented in a development version of the \texttt{DALTON} program.\cite{daltonpaper, dalton2020, olsen2020} 
Furthermore, as a special limit of our range-separated implementation, the non-variational MC-PDFT model CAS-$t$LDA by \citeauthor{li2014multiconfiguration}\cite{li2014multiconfiguration} is included in our implementation:
a CASSCF optimization followed by an energy evaluation with the CAS-sr$t$LDA model (with $\mu=0$) without further optimization of configuration coefficients and orbitals yield the non-variational CAS-$t$LDA model. 
All calculations to be described have been performed in \texttt{DALTON} and the new implementation will be released in a forthcoming release of \texttt{DALTON}.

\ %

\textbf{\ce{H2} potential and bond dissociation energy.} 
The potential energy curve of \ce{H2} was determined using six different methods by varying the interatomic distance on the interval 0.2 Å-4.5 Å with a constant step size of 0.01 Å.
The methods utilized were: full configurational interaction (FCI), CASSCF, non-variational CAS-$t$LDA and the CAS-srDFT utilizing short-range functionals of both LDA-type (srLDA\cite{paziani2006local}), GGA-type (srPBEgws,\cite{goll2005short,goll2006short} referred to as srPBE) and the new translated  sr$t$LDA functional from this work. 
The active space for all CAS methods used consisted of 2 electrons in 2 orbitals ($\sigma_\mathrm{1s}$ and $\sigma^*_\mathrm{1s}$). 
Furthermore, a series of CAS-srDFT (srLDA, srPBE, sr$t$LDA) potential curves with varying values of the range-separation parameter, $\mu$, were performed.
The values attained ranged from 0.2 bohr$^{-1}$ to 1.2 bohr$^{-1}$ with an increment of 0.2 bohr$^{-1}$. 
For all calculations performed on the \ce{H2} molecule, Dunning's aug-cc-pVQZ\cite{dunning1989gaussian,kendall1992electron} basis set was used.

\ %

\textbf{\ce{N2} potential and bond dissociation energy.} A series of energy calculations were performed on the ground-state singlet state ($X ^1 \Sigma^{+}_g$) and the lowest triplet state ($A ^3 \Sigma^{+}_u$)
with a varying interatomic distance in the range 0.4 Å-5.0 Å with an increment of 0.025 Å. For all distances, the energy was determined utilizing the same methods as for the \ce{H2} molecule, for the range-separated methods with $\mu = 1.0$ bohr$^{-1}$.
The CAS space was chosen to comprise of the valence orbitals and electrons ($\sigma_\mathrm{2s}\sigma^{*}_\mathrm{2s}\pi_\mathrm{2p} \sigma_\mathrm{2p}\pi_\mathrm{2p}^{*} \sigma_\mathrm{2p}^{*})^{10}$, i.e., CAS(10,8). 
In the Supplementary Material we provide a comparison of the range-separated singlet potential energy curves for $\mu = 0.4$ bohr$^{-1}$ and $\mu = 1.0$ bohr$^{-1}$.
All calculations were done with Dunning's aug-cc-pVTZ basis set.\cite{dunning1989gaussian,kendall1992electron}

\ %

\textbf{\ce{Cr2} potential energy curves.} The minimum CAS(12,12) active space for correct dissociation was used, corresponding to the 3d and 4s orbitals. 
All calculations were performed with the aug-cc-pVTZ basis set\cite{balabanov2005systematically} and the Douglas-Kroll-He{\ss} (DKH2)\cite{douglas1974quantum,hess1986relativistic,jansen1989revision} second-order scalar relativistic Hamiltonian.
It was verified that the CAS-sr$t$LDA calculations at the dissociation limit corresponded correctly to dissociation into two Cr atoms in their $^7S$ ground states.

\ %

\textbf{Ethene rotational barrier.} To study the rotational barrier of ethene, a minimal CAS(2,2) active space, encompassing the bonding and anti-bonding $\pi$ and $\pi^*$ orbitals, was employed for both the singlet $S_0$ and triplet $T_1$ states.
For the range-separated methods, a range-separation parameter of $\mu=0.4$ a.u.$^{-1}$ was used.
The rotational barrier in ethene was investigated by rotating the dihedral angle spanned by each \ce{H2C} group in an unrelaxed manner from $0^\circ$ to $180^\circ$, starting from the ethene structure described in the benchmark by \citeauthor{loos2018mountaineering}\cite{loos2018mountaineering}.
As reference values for the $S_0$ and $T_1$ states, selected-CI calculations were performed using the QMCPACK software.\cite{kim2018qmcpack}
The calculations were performed on the $0^\circ$ and $90^\circ$ structures for both the $S_0$ and $T_1$ states.
The selected-CI calculations utilized standard settings and a maximum number of determinants of $N_\mathrm{dets}=10^7$.
Calculations on the $T_1$ state was started from a preceding CI singles calculation, whereas the $S_0$ was started from HF orbitals.
All calculations were performed using the aug-cc-pVDZ basis set.\cite{dunning1989gaussian,kendall1992electron}

\section{Results}\label{Results}

\begin{figure*}[htb!]
    \centering
    \includegraphics{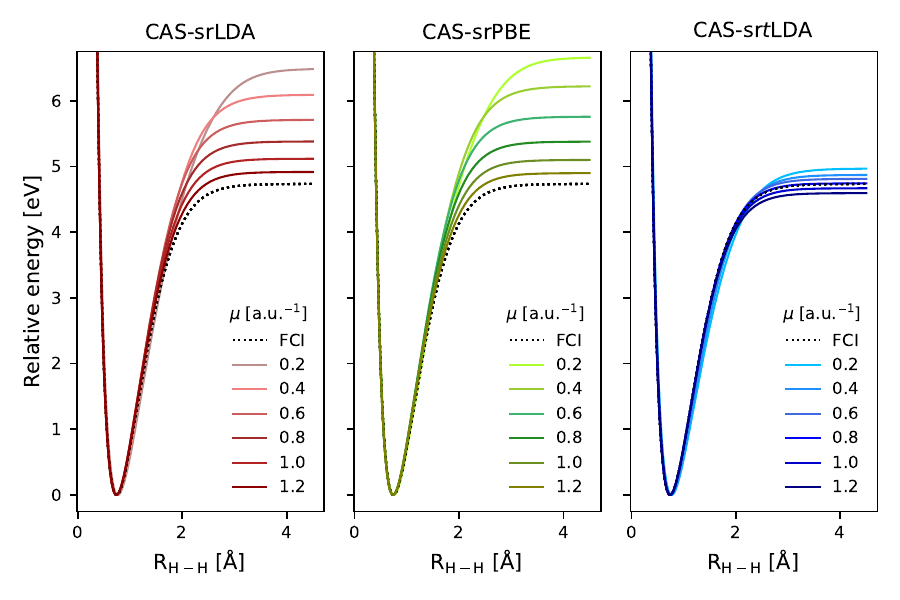}
    \caption{Relative energies of \ce{H2} aligned at the minimum energy calculated for each of the different range separation value ($\mu$) using the traditional CAS-srLDA, CAS-srPBE and the new CAS-sr$t$LDA methods. For all calculations the aug-cc-pVQZ basis set was used.}
    \label{fig:H2_mu_investigation}
\end{figure*}

\begin{figure*}[htb!]
    \includegraphics[scale=0.89]{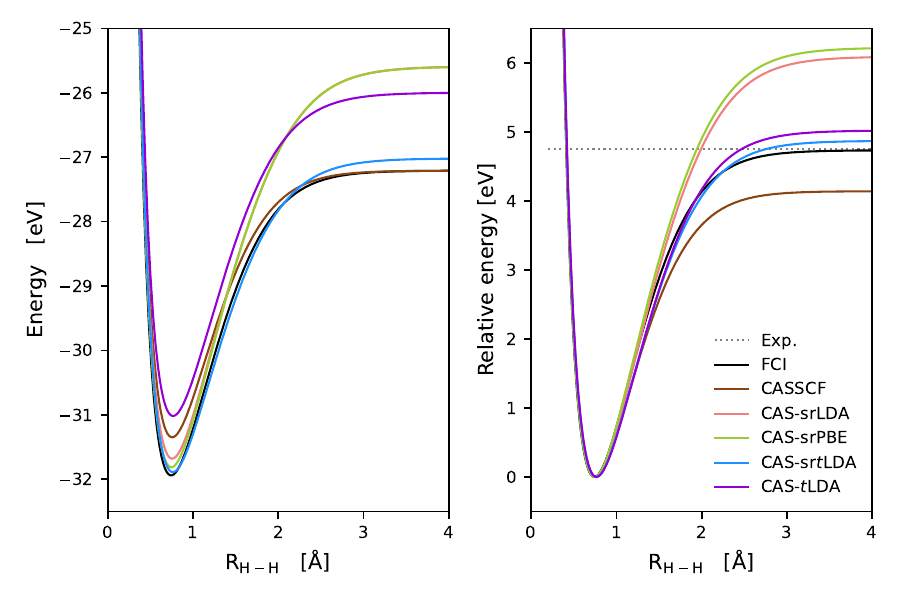}
    \caption{The potential energy curve of $\mathrm{H}_2$ (left) and  potential energy curves aligned at the energy at equilibrium distance for each of the methods employed (right). For all calculations the aug-cc-pVQZ basis set was used. The range-separated hybrid methods all employed $\mu=0.4$ bohr$^{-1}$. The experimental dissociation energy with zero-point corrections ($D_e(\ce{H2})=4.747$ eV) has been taken from Ref. \citenum{herzberg1961dissociation}}
    \label{fig:h2_potential_curve}
\end{figure*}

\textbf{\ce{H2}  dissociation.}  First we investigate how well CAS-sr$t$LDA can describe the ground state potential energy curve of \ce{H2}. In the three frames of Fig.~\ref{fig:H2_mu_investigation} we compare the bond-dissociation energies (BDE) of our previous CAS-srLDA and CAS-srPBE models to the new CAS-sr$t$LDA model for a range of $\mu$ values.
For this simple two-electron system, the singlet spin-restricted CAS-srLDA and CAS-srPBE BDE curves are very similar, and we see that the self-interaction error is growing rapidly when $\mu$ becomes smaller, i.e., when the short-range region is growing towards the full space.
It is gratifying to see the very small $\mu$ dependence of the BDE for the CAS-sr$t$LDA model, clustering around the FCI value.
This shows that the self-interaction error at dissociation is minimized by the on-top pair density here, and in a more robust and smoother way than with unrestricted Kohn-Sham DFT: no unphysical kinks at the Coulson-Fischer point, where the unrestricted solution becomes lower than the spin-restricted solution, are seen for CAS-sr$t$LDA. 
The CAS-sr$t$LDA curves dissociate consistently into two hydrogen atoms described with srLDA, and the remaining small BDE dependency on $\mu$ is caused by the short-range Hartree repulsion not being exactly canceled by the sr$t$LDA exchange and the residual self-correlation in the srLDA functional for a single electron.  A similar conclusion was reached by Hakpa \textit{et al.} with a $\mu$-value of 0.5.\cite{hapka2020a}  
Note that for all three cases one could find a $\mu$ value where the calculated BDE curve is very close to the exact curve; however, this is "the right answer for the wrong reason", a cancellation of errors. 
For growing $\mu$ the missing correlation energy from the long-range CAS wave function grows while the functional error at dissociation becomes smaller (recall that in this special case CASSCF is exact at dissociation), and there is therefore a $\mu$-value where the errors become the same at equilibrium and at dissociation.

In the next figure, Fig.~\ref{fig:h2_potential_curve}, we also compare the three srDFT models at $\mu=0.4$ to other models: FCI, CASSCF, and CAS-$t$LDA. In the left frame we present the absolute calculated energies and in the right frame the relative BDE energies.
On the left one notes that the total energy of the CAS-$t$LDA is the worst, but that is caused by the total energy error in LDA (CAS-$t$PBE performs much better\cite{li2014multiconfiguration}), on the right one sees that the relative energies are a lot better, and only slightly worse than CAS-sr$t$LDA. Because CAS-$t$LDA corresponds to $\mu=0$, it is clear why the residual error is slightly larger than for CAS-sr$t$LDA at $\mu=0.4$, for which the long-range part is exact and not on LDA level.
One also sees that the CAS(2,2)SCF gives too low BDE because of missing dynamic correlation around equilibrium, while CAS-srLDA and CAS-srPBE give much too high BDEs because of the error at large distances discussed above.

\

\begin{figure*}[htb!]
    \centering
    \includegraphics{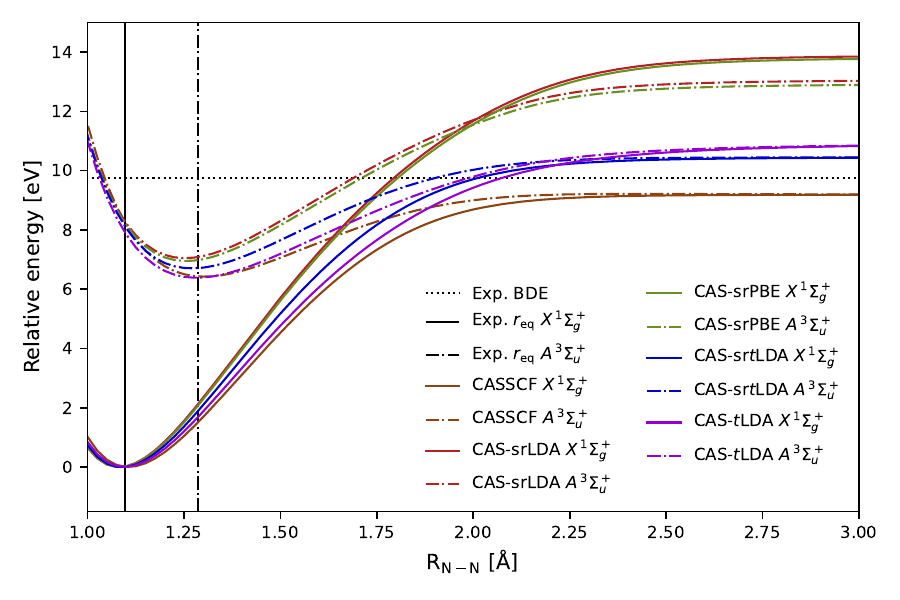}
    \caption{The calculated \ce{N2} singlet state $X\,^1\Sigma_g^+$ (solid line) and triplet state $A\,^3\Sigma_u^+$ (dash dotted line) potential energy curves relative to the singlet minimum energy. For all calculations the aug-cc-pVTZ basis set was used, and all range-separated methods employed a $\mu$-value of $1.0$ bohr$^{-1}$. The vertical solid black and dash dotted lines represent the experimental equilibrium distances of the singlet state (1.09768 Å) and the triplet state (1.2866 Å), respectively.\cite{huber1979molecular}}
    \label{fig:n2_trip_sing}
\end{figure*}

\textbf{\ce{N2}  dissociation.} Compared to  \ce{H2},  the dissociation of the triple bond in \ce{N2} is a more challenging test system. 
For this system, we calculated the potential energy curves of both the ground-state singlet state $(X\,^{1}\Sigma_g^+)$  and the lowest triplet state ($A\,^3\Sigma_u^+$), see Fig. \ref{fig:n2_trip_sing}. It is interesting to compare the potential energy curves of the singlet and triplet states, since they dissociate into the same atomic states containing two spin-coupled nitrogen atoms in their ground state, i.e, \ce{N2}$\rightarrow\ce{N}(^4S) + \ce{N}(^4S)$.
With a correct treatment of the spin-polarization, this degeneracy should thus be predicted at large inter-nuclear distances. 
In order not to overfill the figure, it only contains $\mu = 1.0$ bohr$^{-1}$ curves for all range-separated methods.
In the Supporting Material Figs. S3, S4, and S5 we compare the curves for three $\mu$ values: 0.4, 1.0, and $\infty$ (i.e. CASSCF), similar to Fig. \ref{fig:h2_potential_curve} for \ce{H2}.
As for \ce{H2} above one sees that the singlet spin-restricted CAS-srLDA and CAS-srPBE BDE curves are very similar, and that their self-interaction error is growing rapidly when $\mu$ becomes smaller, i.e., when the short-range region is growing towards the full space.
It is again gratifying to see the small $\mu$ dependence of the BDE for the CAS-sr$t$LDA model, clustering around the experimental value of 9.759 eV.\cite{huber1979molecular}

From Fig. \ref{fig:n2_trip_sing} we see that including the on-top pair density within a translated spin-functional ensures correct degeneracy.
We have verified that all three $M_S$-values of the triplet state gives exactly the same potential energy curve, leading to the correct spin-coupling of the atomic open-shell states with different $M_S$-values. 
The correct degeneracy of different $M_S$ components is not guaranteed for standard approximate DFT functionals, including the functionals used for CAS-srDFT, due to their explicit dependency on the spin-density $\rho_\mathrm{S}$, which is zero everywhere for a singlet state.\cite{becke1995extension,jacob2012spin}
Comparing the singlet and triplet curves produced by the usual srDFT functionals, one notes that they are far from degenerate at dissociation. The triplet curve is lower, presumably because a fraction of the spin-polarization from the unpaired electrons in the atomic state can be described.
However, it is not sufficient for the triplet state to dissociate correctly for CAS-srDFT, the $\rho_\mathrm{S}$ for $M_S = 1$ from CAS-srDFT  can only partly describe the septet atoms at dissociation.  This is again qualitatively the same result as obtained by Hapka \textit{et al.}\cite{hapka2020a} in their non-variational multiconfigurational range-separated on-top pair density hybrid. 
We remark that the energetics at dissociation of the anti-ferromagnetically coupled singlet can also be captured by unrestricted DFT, just as for \ce{H2}, but
that is not the case for the intermediately coupled triplet state.

Experimental values for the equilibrium bond lengths show that the triplet state has a longer equilibrium bond length than the ground-state singlet, as expected from the lowering of the bond order. 
This bond elongation is correctly captured by CASSCF and CAS-sr$t$LDA, whereas CAS-srLDA and CAS-srPBE predict a too short bond.
The curves in Fig. \ref{fig:n2_trip_sing} in fact tell us that the equilibrium distances for CAS-srLDA and CAS-srPBE are too short because the wrong and much too high dissociation limits, in analogy with the textbook example of the restricted Hartree-Fock ground state curve for \ce{H2}.
The CAS-$t$LDA method predicts a BDE in close resemblance with the CAS-sr$t$LDA results which all lie close to the experimental value. 
As CASSCF also here provides a good approximation, one sees exactly the same patterns as for \ce{H2}: CAS-$t$LDA slightly above CAS-sr$t$LDA, CASSCF gives a too low BDE, and CAS-srLDA and CAS-srPBE much too high at long distances. 
Our explanations are the same as for \ce{H2}.
We have also verified, that the CAS-sr$t$LDA value at long distances are indeed two times the CAS(5,4)-sr$t$LDA energy of the ground-state $^4S$ nitrogen atom.

\begin{figure*}[htb!]
    \centering
    \includegraphics{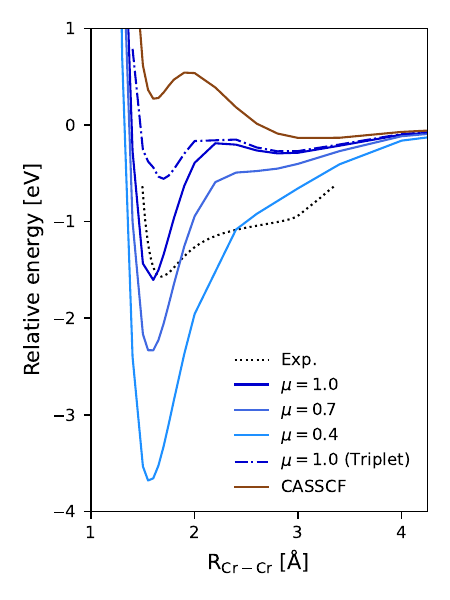}
    \caption{\ce{Cr2} singlet (solid lines) and triplet (dashdot line) CAS-sr$t$LDA potential energy curves aligned at the dissociation energy. The CAS-sr$t$LDA singlet ground state energy curve was calculated at four different $\mu$-values, including CASSCF ($\mu=\infty$), while the triplet curve is only shown for $\mu=1.0$.  All calculations employed the aug-cc-pVTZ basis set and a DKH2 Hamiltonian.
    The experimental curve for the singlet ground state was taken from Ref.\ \citenum{casey1993negative}
    as digitalized in Ref.\ \citenum{larsson2022chromium}.}
    \label{fig:Cr2-fig}
\end{figure*}

\textbf{\ce{Cr2}  dissociation.}  The \ce{Cr2} dimer is a small diatomic molecule where both near-degeneracy effects, spin-coupling, as well as dynamic correlation are crucial to obtain an accurate description of dissociation. 
Therefore, this system presents an interesting challenge for the new MC-sr$t$LDA model.
In its ground-state singlet $X\,^1\Sigma_g^+$ configuration, the \ce{Cr2} dimer forms a formal sextuple bond, which has been experimentally shown to be weak, with an unusually shaped potential energy curve.\cite{maniar2024symmetry}
Experimental data show that the dissociation curve of \ce{Cr2} exhibits a weakly bonding minimum at $1.6788$ Å and forms a "shoulder" at larger distances before asymptotically dissociating into free high-spin $^7S$ \ce{Cr} atoms.\cite{casey1993negative}
This unusual curve shape is attributed to the difference in size and extent of the 3d and 4s orbitals. 
The minimum is primarily caused by 3d-orbital interactions, while the shoulder is due to 4s-4s interactions, with the 3d orbitals coupling anti-ferromagnetically.\cite{larsson2022chromium,muller2009large}

The potential energy curves were determined with  CAS-sr$t$LDA model, using three different $\mu$-values and compared to CASSCF  results as well as the experimental reference. 
The theoretical and experimental potential energy curves are shown in Fig. \ref{fig:Cr2-fig}.
Although an active space of 12 electrons in 12 orbitals encompasses all orbitals responsible for the sextuple bond, Fig. \ref{fig:Cr2-fig} demonstrates that this minimum space is insufficient for a qualitatively correct description of the dissociation when using the CASSCF method. 
Thus, unlike the \ce{H2} and \ce{N2} systems, CASSCF is in this case insufficient to capture the correct features of the dissociation. 
Adding short-range dynamical correlation significantly improves the results and the shape of the potential curve, as evidenced by the different $\mu$-values.
This supports the role of dynamical correlation as a crucial component for a qualitatively correct bonding within the dimer.
However, decreasing the range-separation parameters ($\mu=1.0\rightarrow0.7\rightarrow0.4$) leads to the disappearance of the characteristic shoulder. 

The triplet $A\,^3\Sigma_u^+$ potential curve (Fig. \ref{fig:Cr2-fig}, dashdot line) was determined using CAS-sr$t$LDA to demonstrate once more the accurate prediction of singlet-triplet degeneracy at dissociation, made possible with the on-top pair density functional.
Furthermore, the on-top pair density functional's independence on the $M_S$ value was verified as all $M_S$-values of the triplet where exactly degenerate. 
This is a feature not possible with conventional density functionals due to the high-spin one-electron description of the spin density.
We remark that the energetics at dissociation of the anti-ferromagnetically coupled singlet can also be captured by unrestricted DFT, just as for \ce{H2} and \ce{N2}, but
that is not the case for the intermediately coupled triplet state.

Another significant factor affecting the bonding within \ce{Cr2} is the use of second-order scalar relativistic effects, as introduced by the Douglass-Kroll-He{\ss} approximation. 
Comparisons of the potential energy curves determined with a non-relativistic Hamiltonian and a DKH2 Hamiltonian are shown in the Supplementary Material, Fig. S6.
The significance of the relativistic effects is evident in the drastic deepening of the minimum bond region, while the effects at larger inter-atomic distances are subtle.
The qualitatively correct but quantitatively rather poor description of the potential energy curve as seen when using the CAS-sr$t$LDA can potentially be assigned to the poor performance of the srLDA functional.
Using symmetry-broken DFT-LDA, the quantitative description of potential energy curves have shown to improve significantly upon the addition of semi-local GGA functionals.\cite{patton1997simplified,maniar2024symmetry}
How well one can describe the singlet and triplet curves with a sr$t$GGA functional as sr$t$PBE is thus an interesting case for future investigations.

\ %

\textbf{Ethene rotational barrier.}
To expand the test set utilized in this study to include a heteroatomic example, the rotational barrier of ethene was analyzed for both the singlet $S_0$ ground state and the lowest triplet $T_1$ state. 
All range-separated methods employed a uniform range-separation parameter of $\mu=0.4$ a.u.$^{-1}$, which was identified as the optimal value for organic molecules by \citeauthor{fromager2007universality}\cite{fromager2007universality}. This choice is supported by our investigation of $\mu$-dependence for \ce{H2} and \ce{N2} in this paper, but may require further investigation for functional depending on the on-top pair density. 
\begin{figure}[htb!]
    \centering
    \includegraphics[width=\linewidth]{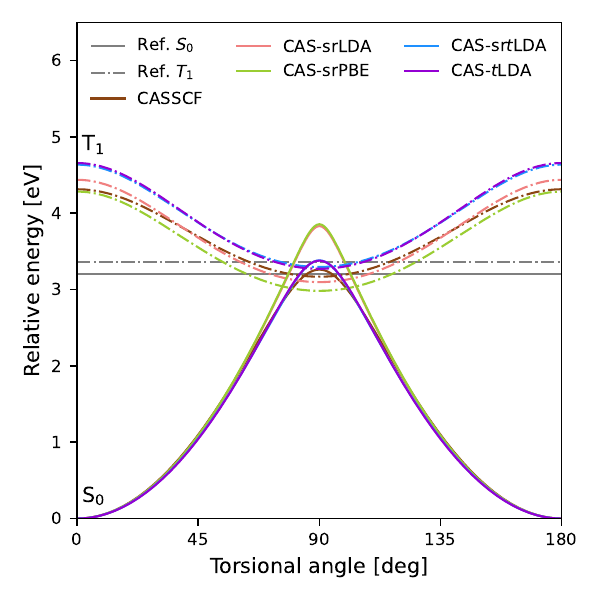}
    \caption{Relative energy of ethene with an increasing torsional angle for the ground-state $S_0$ (solid lines) and the lowest triplet $T_1$ state (dashdot lines). All calculations employ a CAS(2,2) active space and the aug-cc-pVDZ basis set. The range-separated methods employ a $\mu=0.4$ a.u.$^{-1}$. Selected-CI serve as reference values for ethene at $90^\circ$ for both the $T_1$ (grey dashdot line) and $S_0$ (grey solid line).}
    \label{fig:ethene_rotational}
\end{figure}
Examining the potential energy curves of ethene (Fig. \ref{fig:ethene_rotational}), one finds that the differences between the variational CAS-sr$t$LDA and CAS-$t$LDA are negligible.
It indicates that the relaxation of the density and pair-density in CAS-sr$t$LDA is insignificant in this example.
It also shows that range separation is not essential for the description of the rotational barrier, that is, also $\mu=0$ gives good curves here as long as the pair-density is used to describe the local spin density.
The inclusion of the on-top pair density in the short-range exchange-correlation functional significantly decreases the $S_0$-$T_1$ gap at $90^\circ$, while no difference is observed around $0^\circ$ and $180^\circ$.
Comparing the srLDA/srPBE models with the pair density dependent models indicates that the description of the $S_0$ state is prone to self-interaction errors. 
This explains the increasing deviation from both the CASSCF and CAS-sr$t$LDA/$t$LDA model as the mutual stabilization and destabilization of the $\pi$ and $\pi^*$ orbitals increases as the bond is twisted towards the perpendicular conformation.
In contrast, the effect of self-interaction is minimal for the $T_1$ state, as demonstrated by the close agreement between the CASSCF and CAS-srLDA/srPBE potential curves.
When compared to the reference selected-CI values of the $S_0$ state at $90^\circ$, incorporating the pair-density to describe spin polarization significantly improves the quality of the results relative to srLDA/srPBE.
The selected-CI calculations predict the $S_0$ and $T_1$ states to be slightly separated, with the $T_1$ state being higher in energy at $90^\circ$, whereas all other models place the $S_0$ state at a higher energy.
As the CAS(2,2)SCF $S_0$ potential energy deviates only slightly from the reference value, this suggests that the same active space is inadequate to fully describe both states. 
Furthermore, the inclusion of dynamical electron correlation via the exchange-correlation functional appears insufficient to account for the missing effects.

\section{Conclusions}\label{Conclusion}

We have presented the derivation  and implementation of a direct second-order optimization scheme for the new, fully variational MC-srPDFT model based on the charge and on-top pair densities. 
The theory and implementation is generally valid for any short-range functional of $ \rho_\mathrm{C}$ and $\pi$. 
In the calculations presented in this work, the approximate short-range exchange-correlation functional was based on a translation of an existing short-range local spin density approximation functional.
This translation relies on the relationship between the local one-electron spin density and the local two-electron on-top pair density for a single determinant wave function.
This relationship is exact for a single determinant only, and combining it with a multiconfigurational wave function necessitated density screening in practical calculations to avoid regions of space resulting in a complex spin-density.

We have illustrated with CAS-sr$t$LDA how MC-srPDFT models can describe dissociation into the correct atomic states, also for intermediate spin coupling. As expected, the BDEs are significantly improved compared to MC-srDFT models. 
The calculations are also more robust and simpler from a user perspective for non-singlet systems because we obtain the same result for all $M_S$ values.
Compared with the MC-PDFT models introduced by Gagliardi and coworkers (see e.g. the review in Ref.~\citenum{sharma2021multiconfiguration}), the MC-srPDFT is more general in multiple aspects. 
The MC-srPDFT model is fully variational in all wave function parameters.
As shown by Hapka et al.\cite{hapka2020a}, the non-variational combination of a standard CASSCF with PDFT or srPDFT introduces double-counting of the positive kinetic correlation energy, which often causes them to become worse when increasing the active space beyond the minimal for the strong correlation effects, including correct dissociation. This effect is removed with the variational optimization. 
However, the translation of spin-dependent functionals also used in this work causes exaggeration of spin-polarization in bonding regions when used with multi-reference on-top pair densities (as is also the case for unrestricted DFT).\cite{grafenstein2005,hapka2020a}
We have compared CAS(2,2)-sr$t$LDA with CAS(2,13)-sr$t$LDA for \ce{H2} (data not shown) and the BDE error is bigger for the larger CAS, however, it is impossible to say what is caused by srLDA functional error and what is caused by exaggerated spin polarization. We postpone futher analysis of this till we have implemented MC-sr$t$PBE.
Lastly, the use of exact long-range exchange ensure a correct asymptotic behavior of the exchange-correlation potential. 
It should also be noted that, if beneficial, the MC-srPDFT model can straightforwardly be extended to include exact exchange in the short-range functional, e.g. an MC-sr$t$PBE0 model with 25\% exact exchange in the short-range part.

An obvious improvement of our current implementation is to extend the functionals with gradients of the electron density and on-top pair densities since it is generally known that GGA functionals gives equilibrium distances in better agreement with experimental values. Along with improved functionals a benchmark of the method for properties such as excitation energies will also be relevant.\cite{hoyer2016,hapka2020a} 
In this regard, an interesting perspective of the variational construction of the MC-srPDFT with second-order optimization is that the model can straightforwardly be extended to linear response properties, which gives access to frequency-dependent polarizabilites as well as electronic excitation energies and  transition moments (as an alternative to state-average calculations).  
In addition, it is also relatively straighforward to implement analytical molecular gradients, and with more work also molecular Hessians as well as magnetic properties. Work along these lines is in progress. 
Long-range dynamical correlation effects as dispersion, which is not included in MC-srPDFT, can be treated in \texttt{DALTON} with long-range NEVPT2\cite{nevpt2srdft}. 
However, this is computationally expensive, and it would be interesting to implement the simplified long-range adiabatic connection lrAC0 as suggested and tested by Hapka et al.\cite{hapka2020a}
Finally, for even better accuracy the path forward is to construct and test new short-range exchange-correlation functionals explicitly dependent on the on-top pair density from the beginning, removing the single-determinant motivation and the exaggerated spin-polarization of MC wave functions for translated functionals. Work in that direction is also in progress in our group.

\section{Supplementary Material}
In the supplementary material a close-up of the minimum region of the \ce{H2} dimer can be found, detailing the elongation of the \ce{H}--\ce{H} bond when using specific functional types in Fig. S1.
Furthermore, for the same dimer a plot showing the relative energy deviation (relative to FCI) for the CAS-srDFT functional variants with a range of $\mu$-values are provided in Fig. S2, indicating how accurate different points along the potential energy curve are described. 
For the \ce{N2} molecule, in Figs. S3, S4 and S5 the potential energy curves and relative energies (BDE) are compared for two $\mu$ values, 0.4 and 1.0.
The large effect of scalar relativistic corrections (DKH2) to the potential energy curve of \ce{Cr2} is shown in Fig. S6. 


\begin{acknowledgments}
EDH thanks The Villum Foundation, Young Investigator Program (grant no. 29412), the Swedish Research Council (grant no. 2019-04205), and Independent Research Fund Denmark (grant no. 0252-00002B) for support. The computations were performed on computer resources provided by Danish e-infrastructure Cooperation (grant no. DeiC-SDU-L-8) at HIPPO (University of Southern Denmark, SDU).
Furthermore, we thank SDU postdoc Peter Reinholdt for providing the selected-CI reference values of the ethene molecule.

\end{acknowledgments}

\section*{References}

\bibliographystyle{unsrtnat}
\bibliography{references.bib}

\end{document}


\author{Frederik Kamper Jørgensen}
\affiliation{Department of Physics, Chemistry and Pharmacy, University of Southern Denmark, Campusvej~55, DK--5230 Odense M, Denmark}

\author{Erik Rosendahl Kjellgren}
\affiliation{Department of Physics, Chemistry and Pharmacy, University of Southern Denmark, Campusvej~55, DK--5230 Odense M, Denmark}

\author{Hans Jørgen Aagaard Jensen}
\affiliation{Department of Physics, Chemistry and Pharmacy, University of Southern Denmark, Campusvej~55, DK--5230 Odense M, Denmark}
\email{hjj@sdu.dk}

\author{Erik Donovan Hedegård}
\affiliation{Department of Physics, Chemistry and Pharmacy, University of Southern Denmark, Campusvej~55, DK--5230 Odense M, Denmark}
\email{erdh@sdu.dk}

\date{\today}

\title{{\it Supplementary Material: }
\\
Multiconfigurational short-range on-top pair-density functional theory
\\ \ }

\maketitle

\tableofcontents

\clearpage

\section{\ce{H2}: Close-up of equilibrium distance}

\begin{figure*}[htb]
    \centering 
    \includegraphics{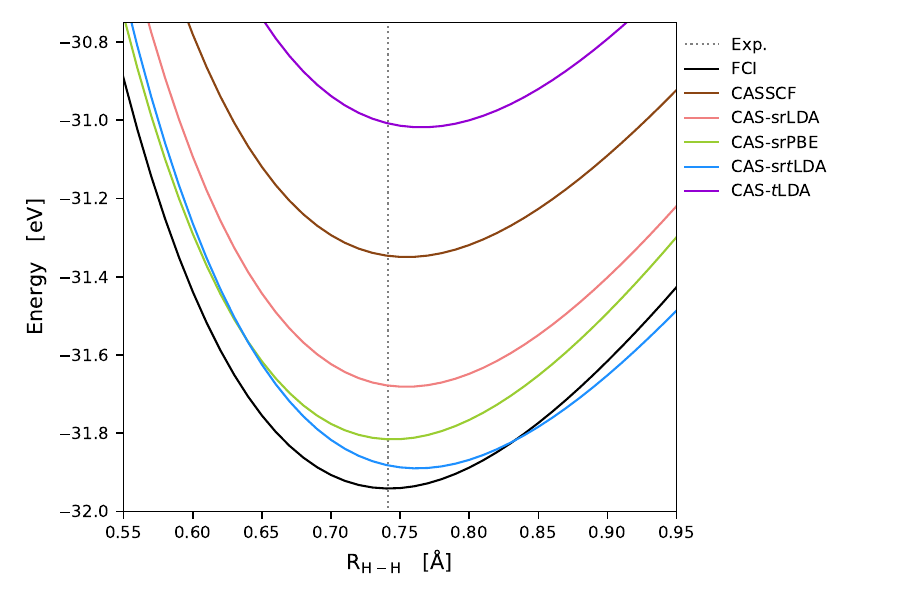}
    \caption{Close-up of the \ce{H2} potential energy curves around equilibrium distance. For all calculations the aug--cc--pVTZ basis set was used with a CAS(2,2) active space. All range--separated methods utilized a range--separation value of $\mu=0.4$ bohr$^{-1}$. The experimental bond length $r_{eq}=0.74144$ Å has been taken from Ref. \citenum{huber1979molecular}.}
    \label{fig:H2_eq_dist}
\end{figure*}

\newpage
\section{\ce{H2}: Relative energy deviations with respect to FCI}
\begin{figure*}[htb!]
    \centering
    \includegraphics{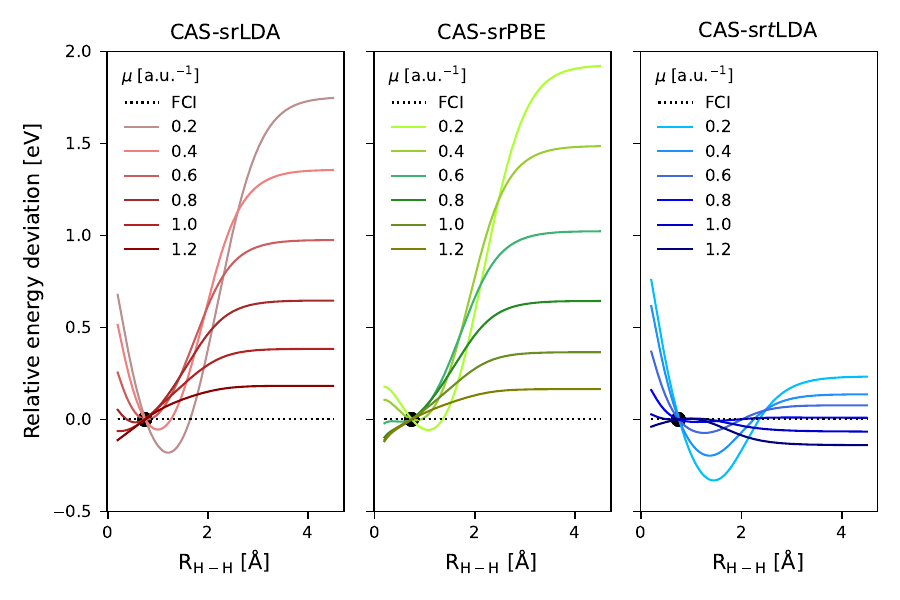}
    \caption{Relative energies using the relative FCI energy as reference. All relative energies have been determined by aligning each potential curve at the minimum energy.}
    \label{fig:mu_relative_to_fci}
\end{figure*}

\newpage

\section{\ce{N2}: Ground-state singlet potential energy curves with range-separation value $\mu=0.4\textrm{ a.u.}^{-1}$}
\begin{figure}[htb!]
    \centering
    \includegraphics[width=\textwidth]{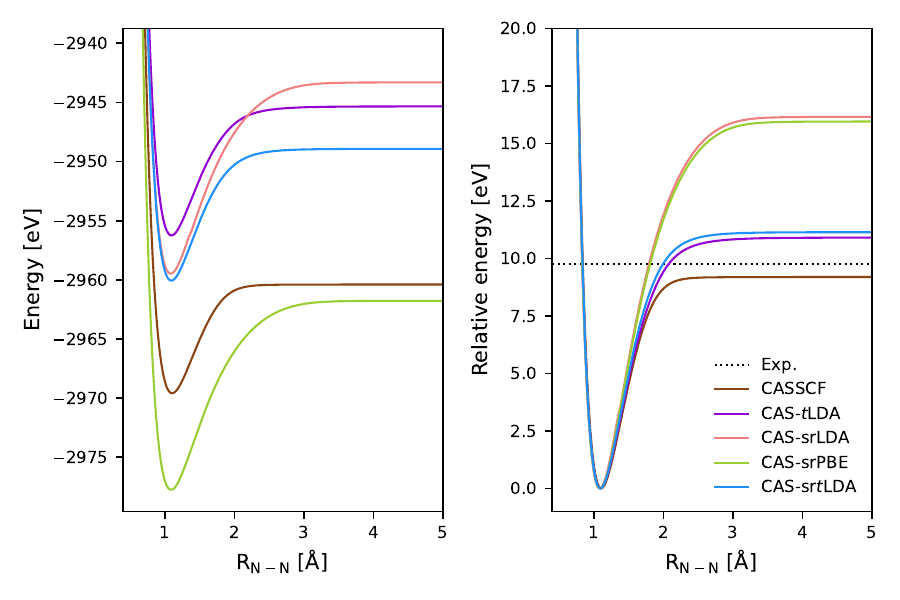}
    \caption{(left) \ce{N2} potential energy curves. (right) \ce{N2} potential curves aligned at the minimum energy. For the range-separated methods, a range-separation value of $\mu=0.4$ a.u.$^{-1}$. All calculations utilized the aug-cc-pVTZ basis set with a full-valence active space of 10 electrons in 8 orbitals.}
    \label{fig:enter-label}
\end{figure}

\newpage
\section{\ce{N2}: Ground-state singlet potential energy curves with range-separation value $\mu=1.0\textrm{ a.u.}^{-1}$}
\begin{figure}[htb!]
    \centering
    \includegraphics[width=\textwidth]{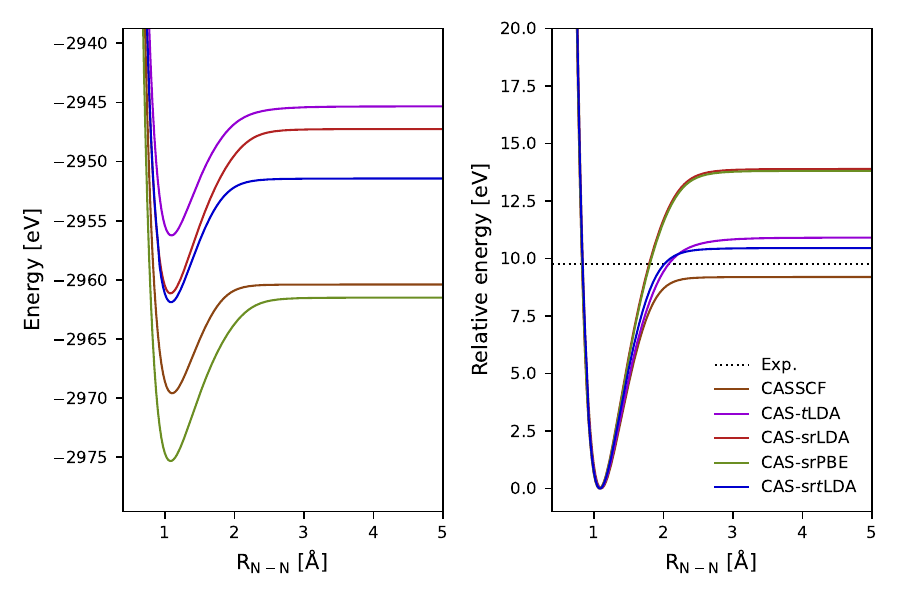}
    \caption{(left) \ce{N2} potential energy curves. (right) \ce{N2} potential curves aligned at the minimum energy. For the range-separated methods, a range-separation value of $\mu=1.0$ a.u.$^{-1}$. All calculations utilized the aug-cc-pVTZ basis set with a full-valence active space of 10 electrons in 8 orbitals.}
    \label{fig:enter-label}
\end{figure}

\newpage
\section{\ce{N2}: Ground-state singlet potential energy curves - comparison of range-separation values}
\begin{figure}[htb!]
    \centering
    \includegraphics[width=\textwidth]{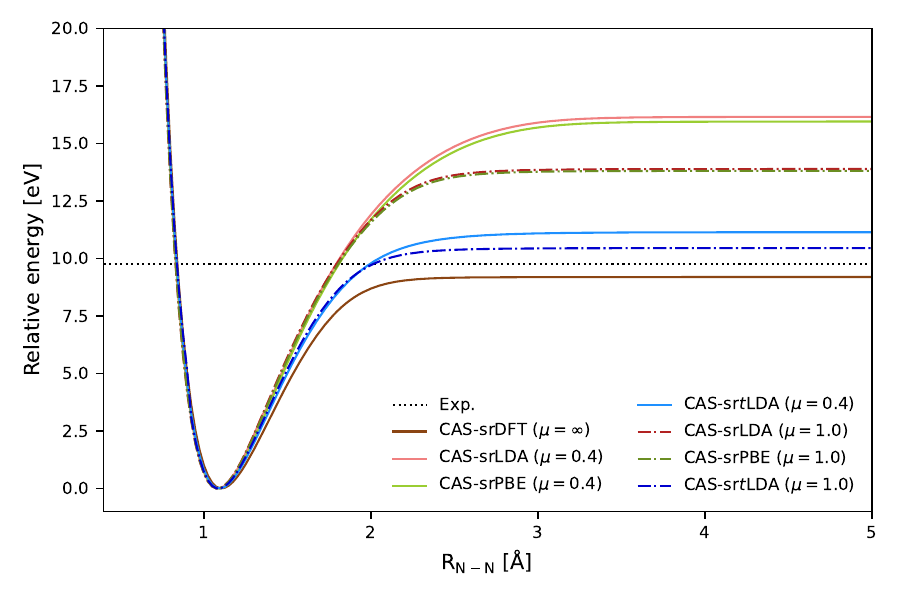}
    \caption{Comparison of the range-separation values: $\mu=0.4$ a.u.$^{-1}$ (solid line), $\mu=1.0$ a.u.$^{-1}$ (dashdot line), and $\mu=\infty$ a.u.$^{-1}$ (solid line). The limit $\mu=\infty$ equals that of the functional independent CASSCF.}
    \label{fig:enter-label}
\end{figure}


\newpage
\section{\ce{Cr2}: Effects of a scalar relativistic Hamiltonian}
\begin{figure*}[htb!]
    \centering
    \includegraphics{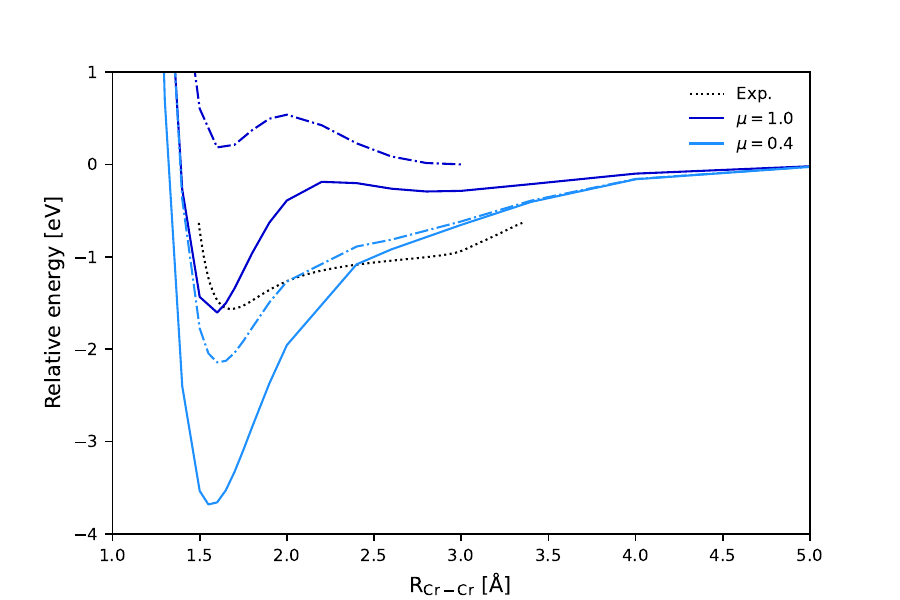}
    \caption{Potential energy curves for the \ce{Cr}--dimer calculated with (solid lines) and without (dashdotted lines) second--order scalar relativistic effects (DKH2). All curves have been calculated using the CAS(12,12)-sr$t$LDA model with the aug-cc-pVTZ basis set and range--separation values of $\mu=0.4$ (light blue) and $\mu=1.0$ (dark blue), respectively. For the experimental curve (dotted line) the data has been obtained in digitalized form by Ref. \citenum{larsson2022chromium} based on the experimental work by \citeauthor{casey1993negative}\cite{casey1993negative}.}
    \label{fig:Cr2-fig-S}
\end{figure*}

\newpage
\bibliographystyle{unsrtnat}
\bibliography{references.bib}